\documentclass[twocolumns, 10pt]{IEEEtran}

\usepackage{amsmath,amssymb}
\usepackage[caption=false,font=footnotesize]{subfig}
\usepackage{graphicx,psfrag,cite,acronym,psfrag}
\usepackage[usenames]{color}
\usepackage{epsfig}
\usepackage{epstopdf}
\usepackage{arydshln}
\usepackage{bm}
\usepackage{url}
\usepackage{soul} 
\usepackage{multirow}

\definecolor{green2}{rgb}{1.0, 0.2, 0.0}
\definecolor{violet}{rgb}{0.54, 0.17, 0.89}

\newcommand{\noteStyleClassicSolid}[1]{\noindent \textcolor{red}{\framebox{$\blacktriangleright$
    {\small\textsf{#1}} $\blacktriangleleft$}}}
\newcommand{\noteStyleClassicWhite}[1]{\noindent \textcolor{red}{\framebox{$\triangleright$
    {\small\textsf{#1}} $\triangleleft$}}}
\newcommand{\noteStylePlain}[1]{\noindent \textcolor{red}{\framebox{\small\textsf{#1}}}}
\newcommand{\noteStyleEmpty}[1]{\marginpar{}}

\newcommand{\mynoteStyleDefault}[1]{\noteStyleClassicSolid{#1}}
\newcommand{\mysubnoteStyleDefault}[1]{\noteStyleClassicWhite{#1}}
\newcommand{\mysubsubnoteStyleDefault}[1]{\noteStylePlain{#1}}

\newcommand{\mapNoteToDefinition}[2]{
\expandafter\renewcommand\csname#1\endcsname[1]{\csname#2\endcsname{##1}}%
}

\newcommand{\setnotelevel}[1]{%
\ifnum#1<1%
    \expandafter\mapNoteToDefinition{mynote}{noteStyleEmpty}%
    \expandafter\mapNoteToDefinition{mysubnote}{noteStyleEmpty}%
    \expandafter\mapNoteToDefinition{mysubsubnote}{noteStyleEmpty}%
\fi%
\ifnum#1=1%
    \expandafter\mapNoteToDefinition{mynote}{mynoteStyleDefault}%
    \expandafter\mapNoteToDefinition{mysubnote}{noteStyleEmpty}%
    \expandafter\mapNoteToDefinition{mysubsubnote}{noteStyleEmpty}%
\fi%
\ifnum#1=2%
    \expandafter\mapNoteToDefinition{mynote}{mynoteStyleDefault}%
    \expandafter\mapNoteToDefinition{mysubnote}{mysubnoteStyleDefault}%
    \expandafter\mapNoteToDefinition{mysubsubnote}{noteStyleEmpty}%
\fi%
\ifnum#1>2%
    \expandafter\mapNoteToDefinition{mynote}{mynoteStyleDefault}%
    \expandafter\mapNoteToDefinition{mysubnote}{mysubnoteStyleDefault}%
    \expandafter\mapNoteToDefinition{mysubsubnote}{mysubsubnoteStyleDefault}%
\fi%
}

\newcommand{\fc} {f_{\text{c}}}

\newcommand{\tp} {\tau_{\text{p}}}

\newcommand{\Tob} {T_{\text{obs}}}

\newcommand{\Ntx}{N_{\text{tx}}}
\newcommand{\Nrx}{N_{\text{rx}}}

\newcommand{\Etot}{E_{\text{tot}}}

\newcommand{\tx}{\text{t}}
\newcommand{\rx}{\text{r}}

\newcommand{\xrm}{x^{\rx}_{m}}
\newcommand{\yrm}{y^{\rx}_{m}}
\newcommand{\zrm}{z^{\rx}_{m}}

\newcommand{\xti}{x^{\tx}_{i}}
\newcommand{\yti}{y^{\tx}_{i}}
\newcommand{\zti}{z^{\tx}_{i}}

\newcommand{\pr}{\mathbf{p}^\rx}
\newcommand{\pt}{\mathbf{p}^\tx}
\newcommand{\prm}{\mathbf{p}^\rx_m}
\newcommand{\pti}{\mathbf{p}^\tx_i}

\newcommand{\DOA}{\bm{\theta}_1}
\newcommand{\doat}{\theta_1}
\newcommand{\doap}{\phi_1}

\newcommand{\aoat}{\theta_1}
\newcommand{\aoap}{\phi_1}
\newcommand{\aoal}{\bm{\theta}_l^{\text{r}}}
\newcommand{\doal}{\bm{\theta}_l^{\text{t}}}
\newcommand{\aoalt}{\theta_l^{\text{r}}}
\newcommand{\aoalp}{\phi_l^{\text{r}}}
\newcommand{\doalt}{\theta_l^{\text{t}}}
\newcommand{\doalp}{\phi_l^{\text{t}}}

\newcommand{\gammalt}{\Delta \theta_l^{\text{r}}}
\newcommand{\gammaltt}{\Delta \theta_l^{\text{t}}}
\newcommand{\gammat}{\Delta \theta^{\text{r}}}
\newcommand{\gammatt}{\Delta \theta^{\text{t}}}
\newcommand{\gammalp}{\Delta \phi_l^{\text{r}}}
\newcommand{\gammalpt}{\Delta \phi_l^{\text{t}}}
\newcommand{\gammap}{\Delta \phi^{\text{r}}}
\newcommand{\gammapt}{\Delta \phi^{\text{t}}}

\newcommand{\orient}{\vartheta^\tx}
\newcommand{\orienp}{\varphi^\tx}
\newcommand{\orien}{\bm{\vartheta}^\tx}
\newcommand{\Thetaot}{\bm{\theta}_0}
\newcommand{\thetaot}{\theta_0}
\newcommand{\phiot}{\phi_0}

\newcommand{\taumr}{\tau_m^{\rx}}
\newcommand{\tauit}{\tau_i^{\tx}}

\newcommand{\taul}{\tau_l}

\newcommand{\alpharl}{\alpha^{\Re}_l}

\newcommand{\alphail}{\alpha^{\Im}_l}

\newcommand{\deltait}{\delta_i^\tx}

\newcommand{\sincro}{\epsilon^\text{s}}
\newcommand{\si}{s_i}
\newcommand{\Si}{S_i}

\newcommand{\pimp}{p_i}

\newcommand{\Pimp}{P_i}

\newcommand{\rxm}{r_m}
\newcommand{\Rxm}{R_m}
\newcommand{\noise}{n_{m}}
\newcommand{\Noise}{N_{m}}

\newcommand{\sumrxm}{\sum_{m=1}^{\Nrx}}

\newcommand{\summpl}{\sum_{l=1}^{L}}

\newcommand{\No}{N_0}
\newcommand{\varsincro}{\sigma_\epsilon^2}

\newcommand{\FIMt}{\mathbf{J}_{\bm{\psi}}}

\newcommand{\FIMtd}{\mathbf{J}^{\text{d}}_{\bm{\psi}}}
\newcommand{\FIMtp}{\mathbf{J}^{\text{p}}_{\bm{\psi}}}

\newcommand{\FIMealfalf}{\mathbf{J}_{\bm{\kappa}\bm{\kappa}}}

\newcommand{\loglik}{\ln f\left(\mathbf{r}, \bm{\psi}_\text{r} \right)}
\newcommand{\loglikcond}{\ln f\left(\mathbf{r} \lvert \bm{\psi}_\text{r} \right)}

\newcommand{\logprio}{\ln f\left(\bm{\psi}_{r} \right)}

\acrodef{Tx}{transmitter}
\acrodef{Rx}{receiver}
\acrodef{POC}{path overlap coefficient}
\acrodef{3D}{three dimensional}
\acrodef{AP}{access point}
\acrodef{AOA}{angle-of-arrival}
\acrodef{D2D}{device-to-device}
\acrodef{MPC}{multipath component}
\acrodef{TDL}{time delay line}

\acrodef{2D}{two dimensional}
\acrodef{OEB}{Orientation Error Bound}
\acrodef{PEB}{Position Error Bound}
\acrodef{PS}{phase shifter}
\acrodef{5G}{fifth generation}

\acrodef{AF}{ambiguity function}

\acrodef{FFT}{fast Fourier transform}
\acrodef{TTD}{true-time delay}

\acrodef{CR}{channel response}

\acrodef{MAP}{maximum a posteriori probability}

\acrodef{MIMO}{multiple-input multiple-output}
\acrodef{mm-wave}{millimeter-wave}

\acrodef{SIMO}{single-input multiple-output}

\acrodef{MISO}{multiple-input single-output}

\acrodef{SISO}{single-input single-output}
\acrodef{TDMA}{time division multiple access}
\acrodef{FDMA}{frequency division multiple access}
\acrodef{CRLB}{Cramer-Rao Lower Bound}

\acrodef{SCAs}{small cell access points}

\acrodef{SCA}{small cell access point}

\acrodef{BS}{base station}

\acrodef{DF}{detect \& forward}

\acrodef{JF}{just forward}

\acrodef{DS}{delay spread}

\acrodef{MLE}{maximum likelihood estimator}
\acrodef{PA}{probability of ambiguities}
\acrodef{CSI}{channel state information}
\acrodef{std}{standard deviation}
\acrodef{RV}{random variable}
\acrodef{SSL}{side-lobe level}
\acrodef{i.i.d.}{independent, identically distributed}

\acrodef{PDF}{probability distribution function}

\acrodef{CDF}{cumulative distribution function}

\acrodef{ch.f.}{characteristic function}

\acrodef{AWGN}{additive white Gaussian noise}

\acrodef{RSSI}{received signal strength indicator}

\acrodef{SNR}{signal-to-noise ratio}

\acrodef{LRT}{likelihood ratio test}

\acrodef{GLRT}{generalized likelihood ratio test}

\acrodef{LOS}{line-of-sight}

\acrodef{NLOS}{non-line-of-sight}

\acrodef{GDOP}{geometric dilution of precision}

\acrodef{GPS}{Global Positioning System}

\acrodef{FIM}{Fisher Information Matrix}

\acrodef{WSN}{Wireless Sensor Network}

\acrodef{MAC}{medium access control}

\acrodef{RSS}{received signal strength}

\acrodef{RTT}{round-trip time}

\acrodef{MF}{matched filter}

\acrodef{ED}{energy detector}

\acrodef{ML}{maximum likelihood}

\acrodef{NL}{nonlinear}

\acrodef{MSE}{mean square error}

\acrodef{RMSE}{root mean square error}

\acrodef{ppm}{part-per-million}

\acrodef{ACK}{acknowledge}

\acrodef{UWB}{ultrawide bandwidth}

\acrodef{TNR}{threshold-to-noise ratio}

\acrodef{NLOS}{non line-of-sight}

\acrodef{LOS}{line-of-sight}

\acrodef{LS}{least squares}

\acrodef{IR-UWB}{impulse radio UWB}

\acrodef{FCC}{Federal Communications Commission}

\acrodef{TH}{time-hopping}

\acrodef{PPM}{pulse position modulation}

\acrodef{PAM}{pulse amplitude modulation}

\acrodef{MUI}{multi-user interference}

\acrodef{PDP}{power delay profile}

\acrodef{BPZF}{band-pass zonal filter}

\acrodef{SIR}{signal-to-interference ratio}

\acrodef{RFID}{radio frequency identification}

\acrodef{WPAN}{wireless personal area networks}

\acrodef{WWLB}{Weiss-Weinstein lower bound}

\acrodef{DP}{direct path}

\acrodef{MF}{matched filter}

\acrodef{MMSE}{minimum-mean-square-error}

\acrodef{SBS}{serial backward search}

\acrodef{NBI}{narrowband interference}

\acrodef{WBI}{wideband interference}

\acrodef{INR}{interference-to-noise ratio}

\acrodef{CIR}{channel impulse response}

\acrodef{LRT}{likelihood ratio test}

\acrodef{RADAR}{RADAR}

\acrodef{MUR}{Multistatic RADAR}

\acrodef{e.m.}{electromagnetic}

\acrodef{CW}{continuous wave}

\acrodef{RF}{radiofrequency}

\acrodef{FCC}{Federal Communications Commission}

\acrodef{EIRP}{effective radiated isotropic power}

\acrodef{RCS}{radar cross section}

\acrodef{BAV}{balanced antipodal Vivaldi}

\acrodef{PRake}{partial Rake}

\acrodef{RTLS}{Real-time locating systems}

\acrodef{EFI}{equivalent Fisher information matrix}

\acrodef{SPEB}{squared position error bound}

\acrodef{SOEB}{squared orientation error bound}

\acrodef{Hi-RADIAL}{High-accuracy RAdio Detection, Identification,
And Localization}

\acrodef{CRB}{Cram\'{e}r-Rao bound}

\acrodef{HCRB}{hybrid Cram\'{e}r-Rao bound}
\acrodef{HFIM}{hybrid Fisher Information Matrix}

\acrodef{ZZB}{Ziv-Zakai bound}

\acrodef{TOA}{time-of-arrival}
\acrodef{DOA}{direction-of-arrival}

\acrodef{ToF}{time-of-flight}

\acrodef{WSN}{wireless sensor network}

\acrodef{MAC}{medium access control}

\acrodef{RSS}{received signal strength}

\acrodef{TDoA}{time difference-of-arrival}

\acrodef{RF}{radiofrequency}

\acrodef{PSD}{power spectral density}

\acrodef{RTT}{round-trip time}

\acrodef{AoA}{angle-of-arrival}

\acrodef{MF}{matched filter}

\acrodef{ED}{energy detector}

\acrodef{ML}{maximum likelihood}

\acrodef{MUR}{Multistatic radar}

\acrodef{HDSA}{high-definition situation-aware}

\acrodef{RRC}{Root raised cosine}

\acrodef{OFDM}{orthogonal frequency division multiplexing}

\acrodef{IF}{intermediate frequency}

\acrodef{PHY}{physical layer}

\acrodef{S-V}{Saleh-Valenzuela}

\acrodef{UHF}{ultra-high frequency}

\acrodef{PR}{pseudo-random}

\acrodef{SoC}{System on Chip}

\acrodef{SoP}{System on Package}

\acrodef{SPMF}{Single-Path Matched Filter}

\acrodef{IMF}{Ideal Matched Filter}

\acrodef{SCR}{signal-to-clutter ratio}

\acrodef{BEP}{bit error probability}

\acrodef{BER}{bit error rate}

\acrodef{WSR}{wireless sensor radar}

\acrodef{HPBW}{half power beam width}

\acrodef{LEO}{localization error outage}

\acrodef{SLAM}{simultaneous localization and mapping}

\begin{document}

\title{Single-Anchor Localization and Orientation Performance Limits using Massive Arrays: \\ MIMO \textit{vs.} Beamforming}

\author{\IEEEauthorblockN{Anna Guerra,~\IEEEmembership{Member,~IEEE,} Francesco Guidi,~\IEEEmembership{Member,~IEEE,}
  ~Davide Dardari~\IEEEmembership{Senior Member,~IEEE}\\}
\IEEEcompsocitemizethanks{\IEEEcompsocthanksitem 

Anna Guerra and Davide Dardari are with 
the Dipartimento di Ingegneria dell'Energia Elettrica e dell'Informazione ``Guglielmo Marconi" - DEI, University of Bologna, via Venezia 52, 47521 Cesena, ITALY. 

(e-mail: anna.guerra3@unibo.it,  davide.dardari@unibo.it).

Francesco Guidi is with CEA, LETI, MINATEC Campus, 38054 Grenoble, France. (e-mail: francesco.guidi@cea.fr).}

\thanks{{\textit{This research was supported in part by the XCycle project (Grant 635975) and by the IF-EF Marie-Curie project MAPS (Grant 659067).
Authors would like to thank Nicol\'o Decarli, Raffaele D'Errico and Henk Wymeersch for fruitful discussions.}}}

}

\maketitle

\begin{abstract}
Next generation cellular networks will experience the combination of  femtocells, \ac{mm-wave} communications and massive antenna arrays. 
Thanks to the beamforming capability as well as the high angular resolution {provided} by massive arrays, only one single \ac{AP} acting as an anchor node could be used for localization estimation, thus avoiding  over-sized  infrastructures dedicated to positioning.
In this context, our paper aims at investigating the localization and orientation performance limits  employing massive arrays both at the \ac{AP} and mobile side. {Thus, we first asymptotically demonstrate the tightness of the \ac{CRB} in massive array regime, and in the presence or not of multipath. Successively,} we propose a comparison between {MIMO} and beamforming  in terms of  array structure, {time} synchronization error and multipath components. Among different array configurations, we consider also {{random weighting}} as a \textit{trade-off} between the high diversity gain of MIMO and the high directivity guaranteed by phased arrays.  {By evaluating the \ac{CRB} for the different array configurations, results show the interplay between diversity and beamforming gain as well as the benefits achievable by varying the number of array elements in terms of localization accuracy.} 
\end{abstract}


\bstctlcite{IEEEexample:BSTcontrol}

\IEEEpeerreviewmaketitle

\section{Introduction}
\label{sec:introduction}
 {The widespread use of personal devices generates new challenges while opening new  appealing scenarios for future applications, such as, for example, those entailing \ac{D2D} interactions or Big Data management issues.} 
To meet these new  {trends}, different disruptive technologies have been recently proposed for the next \ac{5G} wireless communications networks \cite{chin2014emerging, khaitan2011indoor}. In particular, large-scale antenna arrays at \acp{BS} or femtocells \acp{AP} allow to smartly direct the power flux towards intended users thus increasing data rates, whereas \ac{mm-wave} communication provides a less crowded and larger spectrum \cite{larsson2014massive,rusek2013scaling,swindlehurst2014millimeter}. 

 {In next years, it is expected that personal devices localization and communication capabilities will play a crucial role \cite{ di2014location}: in fact, the possibility of localizing nodes in indoor environments will be an essential feature of future devices.} In this context, the \ac{AP} could be used as a single-anchor node, i.e., a node whose position is \textit{a-priori} known, in a radio-localization perspective permitting the mobile users to be aware of their own position. {Furthermore, the adoption of more than one antenna at the {\ac{Tx}} and {\ac{Rx}}, will enable the user orientation estimation at an accuracy higher than that provided by compass and gyroscopes. Such feature could play a key role in applications beyond \ac{5G} as for example augmented reality and \ac{SLAM}, where trajectory errors, comprising both position and orientation estimation inaccuracies, dramatically affect the performance \cite{guidi2016personal}.}
Contrarily to traditional scenarios where dedicated  multiple anchor nodes are necessary to allow classic triangulation/multilateration techniques \cite{dardari2015indoor}, here the possibility to centralize both communication and localization capabilities in a single multi-antenna \ac{AP} working at \ac{mm-wave} frequencies is envisioned with the advantage of drastically decreasing the overall system complexity and cost.  Moreover, when operating at such high frequencies, not only \ac{AP}s but also user terminals could adopt massive arrays  {thanks to the reduced wavelength \cite{hong2014study},} thus increasing even more the localization accuracy given the potential huge set of measurements \cite{razavizadeh2014three,witrisal2016high,guerra2015position,garcia2016direct}.
 
While at microwave frequencies the antenna array technology is quite mature, at \ac{mm-wave} severe technological constraints are still present and must be taken into account when designing positioning systems. Recently, massive antennas prototypes have been proposed with electronic beamsteering capabilities.  In order to reduce the complexity, they adopt simple switches and thus,  {the resulting non-perfect signals phasing operations} could impact the array radiation characteristics \cite{kaouach2011wideband,clemente2013wideband,GuiEtAl:J17}.  {In such a scenario, it becomes of great interest to understand the fundamental limits on localization error with massive antenna arrays both at \ac{AP} and mobile terminals using only a single reference node.} 

 {Concerning the ultimate localization performance evaluation, a rich literature has been produced for the analysis of wideband multiple anchors systems. Specifically, in \cite{shen2010accuracy,shen2010fundamental} authors explore the localization accuracy for a wideband sensors network composed of several independent anchors. Their results are further discussed in \cite{han2016performance} where a more realistic {\ac{Rx} architecture} able to exploit the carrier phases information has been taken in consideration while deriving the localization performance. Differently from these works, where anchors send orthogonal waveforms, we consider a signal model dependent {on} the particular arrays architecture chosen where both orthogonal and non-orthogonal waveforms can be transmitted.  {Moreover, our work is not focused on a specific {\ac{Rx}} structure, as in \cite{han2016performance}, but it aims to {compare} different {{\ac{Tx}} array} architectures.}} In \cite{shahmansoori20155g,mallat2009crbs}, a joint delay-angle estimation is reported considering different array technologies and frequency bandwidths. Nevertheless, these works analyze the performance in terms of \ac{CRB} on  {delay and angular information} rather than  {directly} on localization, and neither a comparison between different array schemes, nor the time synchronization issue and the impact of multipath are treated.  {In our previous work {\cite{guerra2015position,guerra2016position_c}}, some preliminary results on positioning accuracy considering only beamforming strategies have been presented, but the comparison with \ac{MIMO}, as well as the impact of \acp{MPC}, was not considered.}

Stimulated by this framework, in this paper we conduct a \ac{CRB}-based analysis of a localization system exploiting the next \ac{5G} technologies potentialities. 
{Differently from the state-of-the-art, we adopt a 1-step approach in which the {\ac{Tx}} position and orientation are directly inferred from the received signals and, thus, without the need of estimating intermediate parameters (e.g., \ac{TOA}-\ac{DOA}) or applying geometrical methods which do not ensure the optimality of the approach \cite{cover2012elements_c}.} 

{The main contributions of this work can be summarized as follows:
\begin{itemize}
\item  Derivation of the theoretical performance limits on the localization and orientation error for different array configurations in a single-anchor scenario;
\item Proposal of a signal model valid for any antenna array geometry, configuration (i.e., MIMO, phased, timed arrays), and frequency band. As a case study, in the numerical results the focus is on the adoption of mm-wave massive arrays due to their expected attractiveness in next \ac{5G} applications;
\item Introduction of low-complexity random weighting approach, i.e., randomly chosen beamforming weights, and analysis of its performance compared to that of classical beamforming and MIMO solutions;
\item Investigation of the \ac{CRB} tightness in massive array regime (i.e., letting the number of antennas $\rightarrow \infty$) for any \ac{SNR} condition;
\item Analysis of the \textit{trade-off} between \ac{SNR} enhancement obtained via beamforming and diversity gain of \ac{MIMO} considering the impact of different types of uncertainties, as, for example, the \acp{MPC}, beamforming weights and time synchronization errors;
\item Demonstration that in the massive array regime (i.e., array antennas $\rightarrow \infty$), the effect of multipath can be made negligible on average. 
\end{itemize} 
}
{
The rest of the paper is organized as follows. Sec.~\ref{sec:SystemModel} describes the geometry of the localization system. Then, Sec.~\ref{sec:signalmodel} introduces the signal model taking into account different array structures. In Sec.~\ref{sec:posbound} the localization performance limits derivation is reported. Sec.~\ref{sec:crbtight} analyzes the asymptotic conditions for which the \ac{CRB} can be considered a tight bound. Sec.~\ref{sec:idealscenario} derived compact formulas for a ideal free-space case. The multipath impact on localization performance are investigated in Sec.~\ref{sec:mp_loc_ac}.  Finally, Sec.~\ref{sec:numerical} presents the localization performance results and Sec.~\ref{sec:conclusions} concludes the work.
}
\paragraph*{Notation} Lower case and capital letters in bold denote vectors and matrices, respectively. The subscripts $\left[\cdot \right]^{\scriptscriptstyle \text{T}}$, $\left[\cdot \right]^{*}$ and $\left[\cdot \right]^{\scriptscriptstyle \text{H}}$ indicate the transpose, the conjugate and the Hermitian operators. $\lVert \cdot \rVert_2$ is the Euclidean norm, $\mathbf{A} \succeq \mathbf{B}$ indicates that the matrix $\mathbf{A} - \mathbf{B}$ is non-negative definite, and $\text{diag}\left(\cdot\right)$ represents the diagonal operator. The subscripts $(\cdot)^\tx$ and $(\cdot)^\rx$ refer to quantities related to the transmitting and receiving array, respectively, while the subscript $(\cdot)^{\text{tr}}$ to elements that can be referred to both the {\ac{Tx}} and the {\ac{Rx}}. $(\cdot)^\text{FS}$ indicates the free-space scenario. $\mathcal{F}\left(\cdot\right)$ denotes the Fourier transform operation, $\mathcal{U}\left(a,b\right)$ a uniform distribution in the interval $\left[a, b\right]$, and $\mathcal{CN}\left(\mu,\sigma^2\right)$ a circularly symmetric Gaussian distribution with mean $\mu$  and variance $\sigma^2$. 

{The notations of frequently-used symbols are listed as follows.
\begin{description}[\IEEEsetlabelwidth{Very long label}\IEEEusemathlabelsep]
  \item[$\Ntx, \Nrx$ ]   Number of Tx-Rx array antennas 
  \item[$L$ ]            Number of \acp{MPC}
  \item[$A^\tx, A^\rx$ ] Area of the Tx-Rx array 
   \item[$\pt,\orien $ ] Tx centroid position and orientation 
   \item[$\pr,\bm{\vartheta}^\rx$] Rx centroid position and orientation  
   \item[$d$]  Distance between Tx-Rx centroids 
   \item[$S=A^\rx/d^2$]  Ratio between the Rx array area and the squared inter-array distance   
   \item[$\mathbf{p}_i^\tx, \mathbf{p}_m^\rx$] Tx/Rx antenna position    
    \item[$d_\text{ant}$] Inter-antenna spacing   
   \item[$\mathbf{d}\left(\bm{\theta} \right)$] Direction cosine 
\item[$\bm{\kappa}$] Multipath parameters vector 
\item[$\bm{\theta}_1$] Direct path wave  direction    
\item[$\bm{\theta}_l$] $l$th path wave direction    
\item[$\bm{\theta}_0$] Steering direction    
\item[$\tau_{im1}, \tau_{iml}$] Propagation delay relative to the direct and $l$th path between the $i$th Tx-$m$th Rx antenna    
\item[$\tau_1, \tau_l$] Propagation delay relative to the direct and $l$th path between centroids    
\item[$a_1, \alpha_l$ ] Direct path amplitude and $l$th complex channel coefficient
\item[$\tauit (\doal,\orien)$ ] Inter-antenna delay between the $i$th Tx antenna and the relative array centroid 
\item[$\taumr (\aoal,\bm{\vartheta}^\text{r})$ ] Inter-antenna delay between the $m$th Rx antenna and the relative  array centroid 
\item[$\fc, W, \beta$ ] Transmitted signal carrier frequency, bandwidth, baseband effective bandwidth 
\item[$ T_\text{obs}$] Observation interval 
\item[$ E_\text{tot}, E$ ] Total and normalized energy at each antenna element 
\item[$N_0$] Single-side noise power spectral density  
\item[$N_\text{F}$] Receiver noise figure 
\item[$\mathsf{SNR}_1$ ] SNR relative to the direct path
\item[$S_i(f), \mathbf{s}(f)$] Equivalent low-pass signal at the $i$th Tx antenna and transmitted signals vector in the frequency domain
\item[$P_i(f)$] Equivalent low-pass unitary-energy signal at the $i$th Tx antenna in the frequency domain
\item[$R_m(f), \mathbf{r}(f)$] Received signal at the $m$th Rx antenna and received signals vector in the frequency domain
\item[$X_m(f), \mathbf{x}(f)$] Useful Rx signal at the $m$th Rx antenna and  useful Rx signals vector in the frequency domain
\item[$N_m(f), \mathbf{n}(f)$] Noise component at the $m$th Rx antenna and noise vector in the frequency domain
\item[$\omega_i, \mathbf{B}(f, \bm{\theta}_0)$] Beamforming weight and matrix 
\item[$\mu_i^\tx(\bm{\theta}_0), \tau_i^\tx(\bm{\theta}_0), \nu_i$] Beamforming phase, TDL and random weight 
\item[$\delta_i^\tx, \Delta \tau_i^\tx$] Beamforming phase and TDL errors 
\item[$\tilde{\omega}_i, \mathbf{Q}(f)$] Beamforming weight and matrix with errors
\item[$\sincro$] Time synchronization error 
\item[$\bm{\psi}$] Estimation parameter vector
\item[$\mathbf{J}_{\bm{\psi}}, \mathbf{J}_{\bm{\psi}}^{\text{d}}, \mathbf{J}_{\bm{\psi}}^{\text{p}}$] Bayesian FIM, FIM relative to data, \textit{a-priori} FIM 
\item[$\mathsf{CRB}\left(\mathbf{q} \right)$] CRB on position and orientation 
\item[$\mathsf{CRB}_0$] Single-antenna CRB on ranging error 
\end{description}
}

\section{Antenna Array Geometric Configuration}
\label{sec:SystemModel}
\subsection{Geometric Relationships}

\begin{figure}[t!]
\psfrag{tx}[c][c][0.8]{Transmitter \qquad\quad}
\psfrag{rx}[c][c][0.8]{Receiver \quad\quad}
\psfrag{pr}[c][c][0.75]{$\pr$}
\psfrag{prm}[c][c][0.75]{$\,\, \prm$}
\psfrag{pt}[c][c][0.75]{$\pt$}
\psfrag{pti}[c][c][0.75]{$\pti$}
\psfrag{thetarx}[c][c][0.8]{$\theta^{\text{r}}_{m}$}
\psfrag{phirx}[c][c][0.8]{$\phi^{\text{r}}_{m}$}
\psfrag{rrx}[c][c][0.8]{$\rho^{\text{r}}_{m}$}
\psfrag{toa}[c][c][0.8]{$\tau$}
\psfrag{toaim}[c][c][0.8]{\,\,\,$\tau_{im}$}
\psfrag{thetatx}[c][c][0.8]{$\theta^{\text{t}}_{i}$}
\psfrag{phitx}[c][c][0.8]{$\phi^{\text{t}}_{i}$}
\psfrag{rtx}[c][c][0.8]{$\rho^{\text{t}}_{i}$}
\psfrag{Gtx}[c][c][0.65]{$\quad\quad\quad \quad \quad  \!  \text{Tx centroid}$}
\psfrag{Grx}[c][c][0.65]{$\quad\quad\quad \quad \quad \text{Rx centroid}$}
\psfrag{tau}[c][c][0.8]{$\tau^{(1)}$}
\psfrag{tauim}[c][c][0.8]{$\tau_{im}^{(1)}$}
\psfrag{x}[c][c][0.55]{$x'$}
\psfrag{y}[c][c][0.55]{$y'$}
\psfrag{z}[c][c][0.55]{$z'$}
\psfrag{x1}[c][c][0.55]{$x''$}
\psfrag{y1}[c][c][0.55]{$y''$}
\psfrag{z1}[c][c][0.55]{$z''$}
\psfrag{x2}[c][c][0.55]{$x$}
\psfrag{y2}[c][c][0.55]{$y$}
\psfrag{z2}[c][c][0.55]{$z$}
\psfrag{D}[c][c][0.7]{$D$}
\psfrag{appr}[c][c][0.7]{\qquad\qquad\qquad $D \ll c\tau$}
\psfrag{pc}[c][c][0.8]{$\mathbf{p}_\text{0}$}
\psfrag{pant}[c][c][0.8]{$\mathbf{p}_\text{ant}$}
\psfrag{phi}[c][c][0.8]{$\phi$}
\psfrag{th}[c][c][0.8]{$\theta$}
\psfrag{ro}[c][c][0.8]{$\rho$}
\centerline{\includegraphics[width=0.35\textwidth]{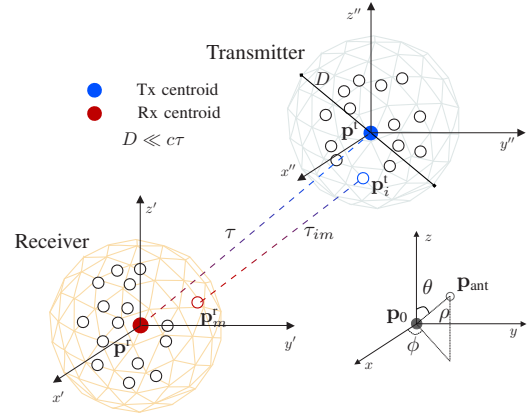}}
\caption{Multi-antenna system configuration.} \label{fig:sys2} 
\end{figure}

We consider a {3D} localization scenario, as the one reported in Fig. \ref{fig:sys2}, consisting of a single \ac{AP} acting as reference receiving node equipped with an antenna array, with $\Nrx$ antennas, and a transmitting mobile terminal with a $\Ntx$-antenna array.
 {The localization process aims at directly inferring:\footnote{ {As previously stated, we consider the {\ac{Tx}} position and orientation with respect to the relative centroid (see \eqref{eq:ch3_coord}-\eqref{eq:Rot_matrix} in the following) as we adopt a 1-step approach in which the {\ac{Tx}} position and orientation are directly inferred from the received signals. Thus, we do not estimate neither the \ac{DOA} (i.e., angle between arrays centroids) nor the direct path \ac{TOA}.} }
\begin{itemize}
\item  the position of the {\ac{Tx}} centroid $\pt~=~\left[x^{\tx}_0, y^{\tx}_0, z^{\tx}_0\right]^{\scriptscriptstyle \text{T}}=\left[x, y, z\right]^{\scriptscriptstyle \text{T}}$;
\item the orientation of the {\ac{Tx}} $\orien=\left[\orient, \orienp \right]^{\scriptscriptstyle \text{T}}$
\end{itemize}}
\noindent when the {\ac{Rx}} centroid position $\pr = \left[x^{\rx}_0, y^{\rx}_0, z^{\rx}_0\right]^{\scriptscriptstyle \text{T}}=\left[0, 0, 0\right]^{\scriptscriptstyle \text{T}}$ and orientation $\bm{\vartheta}^\text{r}=\left[\vartheta^\text{r}, \varphi^\text{r} \right]^{\scriptscriptstyle \text{T}}$ are known.\footnote{Without loss of generality, the {\ac{Rx}} is assumed located at the origin of the coordinates system.} 
With reference to Fig.~\ref{fig:sys2},  $\pti\left(\bm{\vartheta}^\tx \right)= \left[\xti,\,\, \yti,\,\, \zti \right]^{\scriptscriptstyle \text{T}}$ indicates the position  of the $i$th transmitting antenna relative to the {\ac{Tx}} geometric center and dependent on the {\ac{Tx}} orientation, and $\prm(\bm{\vartheta}^\rx)=\left[\xrm,\, \yrm,\, \zrm \right]^{\scriptscriptstyle \text{T}}$ the position  of the $m$th receiving antenna relative to the {\ac{Rx}} geometric center. Considering spherical coordinates, we have
 {
\begin{align}\label{eq:ch3_coord}
&\mathbf{p}_{i/m}^{\text{tr}}(\bm{\vartheta}^{\text{tr}}) \!=\! {\rho}_{i/m}^{\text{tr}}\,\mathbf{R}\left(\bm{\vartheta}^{\text{tr}} \right)\, \mathbf{d}^{\scriptscriptstyle \text{T}}\left(\bm{\theta}_{i/m}^{\text{tr}} \right)
\end{align}
%
with the direction cosine is expressed as
\begin{equation}
\mathbf{d}\left(\bm{\theta} \right)=\left[\sin(\theta)\cos(\phi), \, \sin(\theta)\sin(\phi), \, \cos(\theta)\right]\,
\end{equation}
and ${\rho}_{i/m}^\text{tr}=\lVert \mathbf{p}_{i/m}^{\text{tr}}(\bm{\vartheta}^{\text{tr}})-\mathbf{p}^{\text{tr}} \rVert_2$  and $\bm{\theta}_{i/m}^{\text{tr}}=\left[\theta_{i/m}^{\text{tr}},\phi_{i/m}^{\text{tr}} \right]^{\scriptscriptstyle \text{T}}$ being the distance and the couple of angles between the considered array antenna from the correspondent array centroid.}\footnote{Note that the elevation angle in all the text is indicated with $\theta$ and it can assume values in the interval $\left[0, \pi \right)$. Contrarily the azimuthal angle is denoted with $\phi$ and it ranges between  $\left[0, 2\,\pi \right)$.} 
The rotational matrix $\mathbf{R}(\bm{\vartheta}^{\text{tr}})$ is given by
\begin{equation}\label{eq:Rot_matrix}
\mathbf{R}(\bm{\vartheta}^{\text{tr}})=\mathbf{R}_z(\varphi^{\text{tr}})\, \mathbf{R}_x(\vartheta^{\text{tr}})
\end{equation}
where $\mathbf{R}_z(\varphi^{\text{tr}})$ and $ \mathbf{R}_x(\vartheta^{\text{tr}})$ define the counter-clock wise rotation around the  $z$-axis  and the clock wise rotation around the $x$-axis, respectively. 
Finally $\DOA=\left[ \theta_1, \phi_1 \right]^{\scriptscriptstyle \text{T}}$ designates the angle of incidence between arrays centroids (direct path) and  $\Thetaot=\left[\thetaot, \phiot\right]^{\scriptscriptstyle \text{T}}$ represents the intended pointing direction of the steering process when applied.

The diameter $D$ of the transmitting and receiving arrays is assumed much smaller than the inter-array distance $d=\left\lVert \pr-\pt \right\rVert_2$, i.e., $D \ll d$. Note that this hypothesis is especially  verified at \ac{mm-wave} where the array dimensions are very small  thanks to the reduced wavelength. Moreover the arrays are supposed to be sufficiently far from the surrounding scatterers thus obtaining identical angles of incidence for both direct and \acp{MPC} at each antenna element.

We take $L$ \acp{MPC} into consideration as nuisance parameters in the localization process and the first path is assumed always experiencing a \ac{LOS} propagation condition. For what the \acp{MPC} parameters are concerned, we follow the same notation introduced in \cite{han2016performance}. In particular, let $\doal=\left[ \doalt,\doalp \right]^{\scriptscriptstyle \text{T}}=\left[ \doat+\gammaltt, \doap+\gammalpt\right]^{\scriptscriptstyle \text{T}}$ and $\aoal=\left[ \aoalt,\aoalp \right]^{\scriptscriptstyle \text{T}}=\left[ \aoat+\gammalt, \aoap+\gammalp \right]^{\scriptscriptstyle \text{T}}$, with $l=1,2,\ldots,L$ , indicate the angles of departure from the transmitting array and of incidence at the {\ac{Rx}} side of the $l\,$th path, respectively. The angular biases $\left[\gammaltt, \gammalpt \right]^{\scriptscriptstyle \text{T}}$ and $\left[\gammalt, \gammalp \right]^{\scriptscriptstyle \text{T}}$ are the displacement with respect to the direct path at the {\ac{Tx}} and {\ac{Rx}} side. Obviously, it is $\left[\gammatt_1,\gammapt_1 \right]^{\scriptscriptstyle \text{T}}=\left[ \gammat_1,\gammap_1 \right]^{\scriptscriptstyle \text{T}}=\left[ 0,0 \right]^{\scriptscriptstyle \text{T}}$, when direct path is considered.

Let $\tau_1 \triangleq {\left\lVert \pr - \pt\right\rVert_{2}}/{c}={d}/{c}$ and $\tau_{im1} \triangleq {\left\lVert \prm - \pti\right\rVert_{2}}/{c}$ being the propagation delay related to the direct path between the transmitting and receiving centroids and between the $i$th and $m$th antenna, respectively, where $c$ is the speed of light. Considering the multipath, the $l$th propagation delay between array centroids is defined as $\taul=\tau_1+ \Delta \tau_l$ where $\Delta \tau_l$ is the non-negative delay bias of the $l$th path with $\Delta \tau_1= 0$ \cite{han2016performance}. According to the geometric assumption previously described, the \ac{TOA} and amplitude between each couple of transmitting and receiving antennas can be expressed using the following approximations \cite{mallat2009crbs,han2016performance}
\begin{align}\label{eq:tauapprox}
&1) \,\, \tau_{iml} \approx \taul+\tauit (\doal,\orien)-\taumr (\aoal,\bm{\vartheta}^\text{r}) \quad 2) \,\, a_{iml}\approx a_l
\end{align}
where $a_{iml}$ is the amplitude of the $l$th path between the $m$th receiving  and the $i$th transmitting antenna, and  $\taumr(\aoal,\bm{\vartheta}^\text{r})$ and $\tauit(\doal,\orien)$ are, respectively, the receiving and transmitting inter-antenna propagation delays defined as
\begin{align} \label{eq:ritardi}
&\tau_{i/m}^\text{tr}(\bm{\theta}_l^\text{tr},\bm{\vartheta}^\text{tr})= \frac{1}{c}\,\mathbf{d}\left(\bm{\theta}_l^\text{tr}\right) \, \mathbf{p}_{i/m}^\text{tr} \left(\bm{\vartheta}^\text{tr} \right) 
\end{align}
%
\begin{figure}[t!]
\psfrag{TX}[lc][lc][0.8]{Transmitter}
\psfrag{RX}[lc][lc][0.8]{Receiver}
\psfrag{theta1rx}[lc][lc][0.8]{{$\theta_1^\rx$}}
\psfrag{theta1tx}[lc][lc][0.8]{{$\theta_1^\tx$}}
\psfrag{phi1rx}[lc][lc][0.8]{{$\phi_1^\rx$}}
\psfrag{phi1tx}[lc][lc][0.8]{{$\phi_1^\tx$}}
\psfrag{orient}[lc][lc][0.8]{$\vartheta$}
\psfrag{orienp}[lc][lc][0.8]{$\varphi$\quad}
\psfrag{tauim}[lc][lc][0.8]{$\tau_{im1}$}
\psfrag{taum}[lc][lc][0.8]{$\tau_{m}^\rx$}
\psfrag{taui}[lc][lc][0.8]{$\tau_{i}^\tx$}
\psfrag{theta}[lc][lc][0.8]{$\theta$ \quad}
\psfrag{phi}[lc][lc][0.8]{$\phi$}
\psfrag{x}[lc][lc][0.7]{$x$}
\psfrag{y}[lc][lc][0.7]{$y$}
\psfrag{z}[lc][lc][0.7]{$z$}
\psfrag{tau}[lc][lc][0.7]{$\tau_1$}
\psfrag{tau2}[lc][lc][0.7]{$\tau_2$}
\psfrag{tau2mez}[lc][c][0.7]{$\tau_2 / 2$}
\psfrag{i}[lc][lc][0.6]{$i$th}
\psfrag{m}[lc][lc][0.6]{$m$th}
\psfrag{ptx}[lc][c][0.6]{$\pt$}
\psfrag{prx}[lc][c][0.6]{$\pr$}
\psfrag{a}[lc][lc][0.6]{(a)}
\psfrag{b}[lc][lc][0.6]{(b)}
\psfrag{c}[lc][lc][0.6]{(c)}
\psfrag{dant}[lc][lc][0.7]{$d_\text{ant}$}
\psfrag{m}[lc][lc][0.7]{$m$th}
\psfrag{mx}[lc][lc][0.7]{$m_x\,d_\text{ant}$}
\psfrag{mz}[lc][lc][0.7]{$m_z\,d_\text{ant}$}
\psfrag{x}[lc][lc][0.7]{$x$}
\psfrag{z}[lc][lc][0.7]{$z$}
\psfrag{a1}[lc][lc][0.7]{$\frac{\sqrt{N}}{2}$}
\psfrag{b1}[lc][lc][0.7]{$-\frac{\sqrt{N}}{2}$\quad }
\centerline{\includegraphics[width=0.5\textwidth]{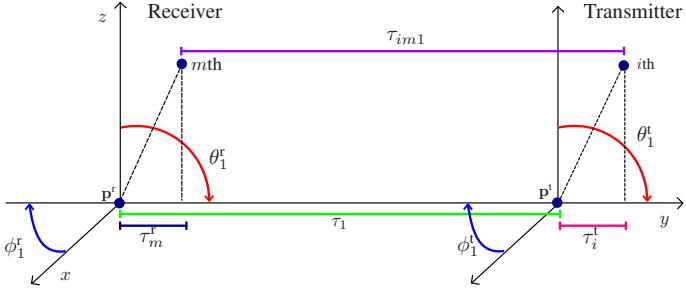}}
\caption{Array geometric configuration.} \label{fig:fig1} 
\end{figure}
Fig.~\ref{fig:fig1} reports a graphical explanation of the considered system and delays. As it can be seen,  the approximation in \eqref{eq:tauapprox} permits to write the \ac{TOA} as the summation of the inter-antenna delays and  the delay between the array centroids.
\begin{figure*}[t]
  \begin{minipage}{0.95\textwidth}
\psfrag{label0}[c][c][0.8]{$S(f)$}
\psfrag{label1}[c][c][0.8]{$\tau_1^\text{t}(\bm{\theta}_0)+\Delta \tau_1^\tx$}
\psfrag{label2}[c][c][0.8]{ {$\mu_1^\tx(\bm{\theta}_0)+\delta_1^\tx$}}
\psfrag{label2r}[c][c][0.8]{$ {\upsilon_1}$}
\psfrag{label3}[c][c][0.8]{$\tauit(\bm{\theta}_0)+\Delta \tau_i^\tx$}
\psfrag{label4}[c][c][0.8]{ {$\mu_i^\tx(\bm{\theta}_0)+\delta_i^\tx$}}
\psfrag{label4r}[c][c][0.8]{$ {\upsilon_i}$}
\psfrag{label5}[c][c][0.8]{$\tau_{\Ntx}^\text{t}(\bm{\theta}_0)+\Delta \tau_{\Ntx}^\tx$}
\psfrag{label6}[c][c][0.8]{ {$\mu_{\Ntx}^\tx(\bm{\theta}_0)+\delta_{\Ntx}^\tx$}}
\psfrag{label6r}[c][c][0.8]{$ {\upsilon_{\Ntx}}$}
\psfrag{label7}[c][c][0.8]{$S_{1}(f)$}
\psfrag{label8}[c][c][0.8]{$S_{i}(f)$}
\psfrag{label9}[c][c][0.8]{$S_{\Ntx}(f)$}
\psfrag{TDL}[c][c][0.65]{$\mathsf{TDL}$}
\psfrag{timed}[c][c][0.8]{(b) Timed Array}
\psfrag{phas}[c][c][0.8]{(a) Phased Array}
\psfrag{mimo}[c][c][0.8]{(c) MIMO}
\psfrag{rand}[c][c][0.8]{(d) {Random Weighting}}
\centerline{\includegraphics[width=0.85\textwidth]{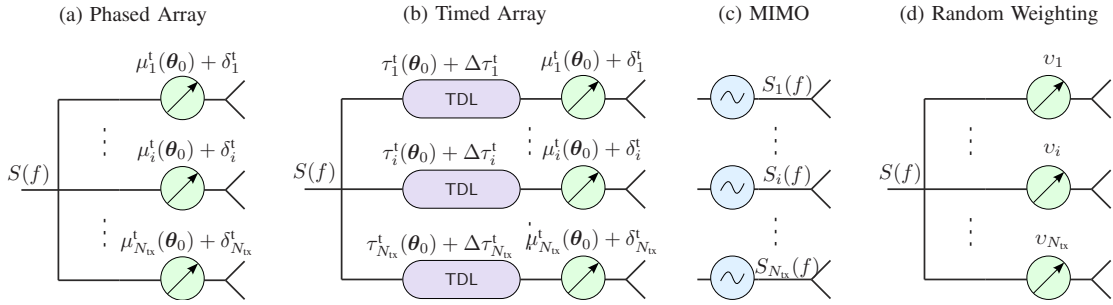}}
\caption{From the left to the right: Phased, timed, MIMO and  {{random weighting}} array schemes.} \label{fig:musch} 
  \end{minipage} 
\end{figure*}
\subsection{ {Special Case: Planar Array Geometry}}
\label{sec:planar}
Planar array configurations appear to be the most suitable when considering the integration of massive arrays in portable devices or in small spaces.  {For this reason, in addition to the general analysis valid for any geometric configuration (i.e., any antennas spatial deployment and arrays orientation),} some compact specialized equations will be derived in the following sections for squared arrays of area $A^\tx\!=\!d^2_\text{ant} {\Ntx}$ ($A^\rx\!=\!d^2_\text{ant} {\Nrx}$), with the antennas equally spaced of $d_\text{ant}$. Both arrays are considered lying on the $XZ$-plane and being located one in front of the other with  $\pr\!=\!\left[0,\, 0,\, 0 \right]^{\scriptscriptstyle \text{T}}$ and $\pt=\left[0,\, y,\, 0 \right]^{\scriptscriptstyle \text{T}}$ with $y>0$, so that $d\!=\!y$ and thus $\tau_1=y/c$. In this case, the antenna coordinates in  \eqref{eq:ch3_coord} becomes
\begin{align}\label{eq:ch3_coord_planar}
\!\!\!\!\!\mathbf{p}_m^\text{tr} \left(\bm{\vartheta}^\text{rt} \right)\!=\! \left[x_m^\text{tr},\, 0,\, z_m^\text{tr} \right]^{\scriptscriptstyle \text{T}}\!\!=\!\mathbf{R}\left(\bm{\vartheta}^\text{tr}\right)  \! \left[m_x\, d_\text{ant},\, 0,\, m_z\, d_\text{ant} \right]^{\scriptscriptstyle \text{T}}  
\end{align}
where $m_x=m_z=-\frac{\sqrt{N}-1 }{2},-\frac{\sqrt{N}-1}{2}+1, \ldots, \frac{\sqrt{N}-1}{2}$ are the antenna indexes along the $x-$ and $z-$axis respectively, and $N$ indicates the number of antennas.

We assume for now a free-space propagation condition so that $\bm{\theta}_1^\tx\!=\!\bm{\theta}_1^\rx\!\!=\!\!\bm{\theta}_1\!\!= {\left[\theta_1,\phi_1 \right]^{\scriptscriptstyle \text{T}}} =\!\!\left[\frac{\pi}{2},-\frac{\pi}{2} \right]^{\scriptscriptstyle \text{T}}$ and $\mathbf{d}(\bm{\theta}_1)=\left[0, \, -1, \, 0\right]$. Consequently it is possible to specialize \eqref{eq:ritardi} as
\begin{align}\label{eq:ritardi_planar}
&\taumr(\bm{\theta}_1,\bm{\vartheta}^\text{r})= -\frac{d_\text{ant}}{c}\,\left(m_x\, \sin\left(\varphi^\rx\right)+m_z \, \cos\left(\varphi^\rx \right)\, \sin\left(\vartheta^\rx \right) \right) \nonumber \\
&\tauit(\bm{\theta}_1,\orien)=  -\frac{d_\text{ant}}{c}\,\left(i_x\, \sin\left(\varphi^\tx\right)+i_z \, \cos\left(\varphi^\tx \right)\, \sin\left(\vartheta^\tx \right) \right)
\end{align}
%
Note that,  {in the special case in which} the {\ac{Rx}} and {\ac{Tx}} orientation is $\bm{\vartheta}^\text{r}=\bm{\vartheta}^\text{t}=\left[0, 0 \right]^{\scriptscriptstyle \text{T}}$, the inter-antenna delays are zeros, i.e. $\taumr(\bm{\theta}_1,\bm{\vartheta}^\text{r})=\tauit(\bm{\theta}_1,\orien)=0$ $\forall m,i$, as the antennas are aligned to the array centroids, thus the incident wave impinges simultaneously at all the antennas.

\section{Antenna Array Schemes and Signal Model}
\label{sec:signalmodel}
In this section, different types of antenna array schemes are analyzed starting from a unified signal model with the purpose to highlight their beamforming and diversity gain properties.  Specifically, the four array structures reported in Fig.~\ref{fig:musch} will be analysed from a signal processing point-of-view and by focusing on how the different signaling schemes translate into different localization capabilities. {Table \ref{tab:comparison_array} reports a comparison in terms of arrays complexity, capabilities and cost.}
%
%
\begin{table*}\caption{{Array schemes comparison.}}
\label{tab:comparison_array}
\begin{minipage}{0.95\textwidth}
\begin{center}%
{
\begin{tabular}{|l l l c c|}
\hline  &&&& \\
 & Signal Design  & Array  Implementation Complexity & Cost & Beamforming capabilities  \\ &&&& \\ \hline  &&&& \\
Timed  & Low: same signal for all antenna branches &  High: {TDLs} needed when $W\gg \fc$  &  High  &  Yes \\ &&&& \\
Phased   & Low: same signal for all antenna branches  & Medium: only {PSs} & Medium &  Yes   \\ &&&& \\
MIMO  &  High: one different signal for all antenna branches  &  High: a RF chain for each branch & High  &  No \\  &&&& \\
Random  & Low: same signal for all antenna branches   & Low: only {PSs} & Low&No\\ 
&  &    &  &  \\
\hline
\end{tabular}
}
\end{center}
\end{minipage}
\end{table*}
%
\subsection{Transmitted Signal Model}
\label{sec:ch3_txmodel}
The transmitted signal at the $i$th transmitting antenna is denoted with $g_i(t)=\,\Re\left\{ \si (t) \, e^{j\, 2\, \pi \fc t}\right\}$ where $\si (t) $ represents the equivalent low-pass signal and $\fc$ the carrier frequency.
We consider a constraint on the total transmitted energy $\Etot$ which is uniformly allocated 
among antennas, thus $E \!\!=\!\! \Etot / \Ntx \!\!=\!\! \int_{\Tob} \lvert \si(t) \rvert^2 dt$, $i=1,2,\ldots, \Ntx$, represents the normalized energy at each antenna element. We introduce the Fourier transform of $\si(t)$ as $\Si(f)=\mathcal{F}\left\{ \si(t) \right\}$, with $\mathcal{F}\left\{\cdot \right\}$ denoting the Fourier transform operation in a suitable observation interval $\Tob$  containing the signal support. For further convenience, the vector $\mathbf{s}(f)=\left[S_1(f),\, \ldots,\, S_{\Ntx}(f) \right]^{\scriptscriptstyle \text{T}}$ contains all the baseband {transmitted} signals. In the following, the signal model for each array configuration will be further detailed with reference to Fig.~\ref{fig:musch}.
\subsubsection{Timed and phased arrays}
\label{sec:ch3_txtimed}
In multi-antenna systems, beamforming is obtained by applying a progressive time delay at each array element so that the emitted signals result summed up coherently towards the intended steering direction.   {Considering the signal bandwidth $W$, when the condition $W \ll \fc$ holds,} this process can be well approximated using only \acp{PS} (phased arrays). On the contrary, when $W \approx \fc$, phase shifts are no longer sufficient to align all the signals. As a consequence, to avoid distortion and beamsteering\footnote{The terms beamsteering and beamforming are used as synonymous.} 
degradation (squinting effect), timed arrays consisting of \acp{PS} and \acp{TDL} must be introduced. {The following analysis considers both array structures in order to preserve the generality of the model. Nevertheless, in Sec.~\ref{sec:numerical}, arrays operating at \ac{mm-wave} frequencies with $W \ll \fc$ (narrowband) will be adopted in simulations.\footnote{{As expected, since $W \ll \fc$, the localization performance of timed and phased arrays coincides.}}
In \cite{guerra2015position}, different fractional bandwidths have been taken into account in the results.}

{Moreover, differently from \cite{alkhateeb2014channel_c,garcia2016location_c}, where multiple beams are generated, here we consider a single-beam scenario in order to maximize the \ac{SNR} in the desired steering direction and to reduce the processing time.}

Given these array schemes, the transmitted signal is the same for all transmitting antennas, i.e., ${s}_i(t)={s}(t)=\sqrt{E}\,\, p(t) \,\,\, \forall i=1,\, \ldots,\, \Ntx$, with $p(t)$ being the unitary energy normalized version of $s(t)$, and beamforming weights are applied to each branch of the array to focus the power flux in a precise direction in space. Specifically, when no quantization errors are present in the weights, the ideal beamforming matrix can be defined as
\begin{align}\label{eq:beamformvect}
&\mathbf{B}\left(f, \bm{\theta}_0 \right)=\text{diag}\left(\omega_1,\, \omega_2,\,\ldots,\, \omega_i,\, \ldots\, \omega_{\Ntx} \right).
\end{align}
\newcounter{MYtempeqncnt0}
\begin{figure*}[t!]
\normalsize
\setcounter{MYtempeqncnt0}{\value{equation}}
\begin{align}\label{eq:rxsignal}
&\mathbf{r}(f)=\summpl \mathbf{a}^{\rx}(f,  \aoal,\bm{\vartheta}^\text{r})\, \mathbf{c}(f,\taul)\, \mathbf{A}^{\tx}(f, \doal, \orien)\, \mathbf{Q}(f)\, \mathbf{B}(f, \Thetaot)\, \mathbf{s}(f)+ \mathbf{n}(f) =\mathbf{x}(f)+\mathbf{n}(f) 
\end{align}
\hrulefill
\vspace*{4pt}
\end{figure*} 
The $i$th beamforming weight is $\omega_i \!\!=\!\! b_i(f)\, b_i^\text{c}$ having indicated $b_i(f) \!\!=\!\! e^{j 2\, \pi\, f \, \tau^{\tx}_{i}(\Thetaot)}$ and  {$b_i^\text{c}\!\!=\!\! e^{j \mu_{i}^{\tx}(\Thetaot)}$}, where  {$\mu_i^\tx(\Thetaot) \!\!=\!\! 2\, \pi\, \fc\, \tau^{\tx}_{i}(\Thetaot)$} and $\tau^{\tx}_{i}(\Thetaot)$ are the transmitting steering phase and delay related to the $i$th \ac{PS} and \ac{TDL} of the array, respectively. The main difference between phased and timed array is the way in which the beamsteering process is performed: in the former only \acp{PS} are present (i.e.,  $\tau^{\tx}_{i}(\Thetaot)\!\!=\!\!0$  $\forall i=1,\,\ldots,\, \Ntx$, refer to Fig.~\ref{fig:musch}-(a)) while in the latter \acp{TDL} and \acp{PS} are both employed to counteract the beamsquinting effect caused by a larger $W/\fc$ ratio (see Fig.~\ref{fig:musch}-(b)). Nevertheless, some technological issues could induce errors in the beamforming vector.  Firstly, when digitally controlled \acp{PS} are used in place of their high-resolution analog counterparts, the presence of quantization errors has to be accounted for \cite{guidi2016personal}. {As shown in \cite{kaouach2011wideband,clemente2013wideband}, where some massive arrays prototypes  working in the X- and V-bands have been proposed, \acp{PS} can be realized by simply adopting switches, or by rotating patch antennas. Therefore, continuous phase shifts ranging from $0^\circ$ to $360^\circ$ are not realizable in practice and the quantization errors generated by the consequent discretization of phases should be taken into account when considering real massive arrays.} Secondly,  {time synchronization} between the {\ac{Tx}} and the {\ac{Rx}} is required to estimate the position. There are several techniques to accomplish this task \cite{dardari2009ranging}, with the two-way ranging being one of the most used. Unfortunately due to several factors such as clock drift, a residual {time synchronization} error is always present and it is accounted by the term $\sincro$ in our model.  In the presence of such non-perfect weights and {time synchronization} error, a matrix accounting for all the non-idealities is introduced
\begin{align}\label{eq:errorvect}
\!\!\!\!&\mathbf{Q}(f)= e^{-j\,2\, \pi\, \left(f+\fc\right) \sincro}\,\text{diag}\left(\varsigma_1,\, \varsigma_2,\, \ldots,\, \varsigma_i,\, \ldots,\, \varsigma_{\Ntx}\right)
\end{align}
where $\varsigma_i$ takes into account the $i\,$th beamforming weight quantization error, i.e., $\varsigma_i=e^{j  \left(2\, \pi\, f \, \Delta \tau^{\tx}_i+\delta_i^{\tx}  \right)}$ with $\deltait$ being the phase error and  $\Delta \tauit$ the \ac{TDL} error.  For further convenience, let indicate with {$\tilde{\omega}_i=\tilde{b}_i(f)\, \tilde{b}_i^{\text{c}}$ where} $\tilde{b}_i(f) \!\!=\!\! e^{j 2\, \pi\, f \,( \tau^{\tx}_{i}(\Thetaot)+ \Delta \tau^{\tx}_i)}$ and  {$\tilde{b}_i^{\text{c}}\!\!=\!\! e^{j (\mu_{i}^{\tx}(\Thetaot)+\delta_i^{\tx} )}$} the quantized weights. After the transmitting beamforming process, the signal at each antenna element can be written as  $\mathbf{Q}(f)\, \mathbf{B}(f, \Thetaot)\, \mathbf{s}(f)$. 

\subsubsection{MIMO arrays}
\label{sec:ch3_txmimo}

Contrarily to timed or phased arrays, which perform accurate beamforming, \ac{MIMO} arrays take advantage of  {the diversity gain provided} by multiple {different} waveforms \cite{li2007mimo,fishler2006spatial} (see Fig.~\ref{fig:musch}-(c)).\footnote{ {Note that here we refer to \ac{MIMO} as done in radar literature rather than in communications.}} To make the {\ac{Rx}} able to discriminate the signal components coming from each single transmitting antenna, orthogonal waveforms are typically adopted \cite{hassanien2010phased,fishler2006spatial,he2010target,haimovich2008mimo}.  As an example, in \cite{fishler2006spatial} a class of signals (i.e., frequency spread signals) are demonstrated to maintain orthogonality for time delays and frequency Doppler shifts. This comes at the expense of a large bandwidth or symbol duration time and of a higher complexity. In \ac{MIMO} arrays, the normalized baseband transmitted signals are indicated with $\Pimp(f)=\mathcal{F}\left\{ \pimp(t) \right\}=\frac{1}{\sqrt{E}}\, \mathcal{F}\left\{\si (t)\right\}$, where $\int_{\Tob} \lvert \pimp(t) \rvert^2 dt=1, i=1,2, \ldots, \Ntx$.  We consider orthogonal waveforms, such that the correlation function is
\begin{align}\label{eq:orth}
R_p\left(\Delta \tau_{ij}^{(l,k)} \right) & = \! \int_{W} \, P_{i}(f)\,P_{j}^{*}(f)\, e^{-j 2 \pi f \Delta \tau_{ij}^{(l,k)}}\, df \nonumber \\
&=\begin{cases}
& \!\! 0 \qquad \,\,\,\,\,\, \text{$i\neq j$}  \\
&\! \! \neq 0 \qquad \text{$i=j$}
\end{cases} \quad {\forall l,k=1,\ldots,L}
\end{align}
{where $\Delta \tau_{ij}^{(l,k)}=\tau_{iml}-\tau_{jmk}$ with $m=1,\ldots, \Nrx$ and $i,j=1,\ldots, \Ntx$}. The possibility to provide orthogonal waveforms permits to increase the diversity gain, as it will be detailed in next sections, but it requires a greater bandwidth demand and  a more complex {\ac{Tx}} structure. In \ac{MIMO}, the matrix in \eqref{eq:beamformvect} is an identity matrix $\mathbf{B}\left(f,  \bm{\theta}_0 \right)=\mathbf{B}=\mathbf{I}_{\Ntx}$.  In presence of the {time synchronization} error, \eqref{eq:errorvect} becomes $\mathbf{Q}(f)= e^{-j\,2\, \pi\, \left(f+\fc\right) \sincro}\,\mathbf{I}_{\Ntx}$. 

\subsubsection{{Random  Weighting}}
\label{sec:ch3_txrandom}
To avoid the complexity of MIMO, we propose a strategy relying on the same structure of phased arrays, i.e., with only \acp{PS} at each antenna branch (see Fig.~\ref{fig:musch}-(d)), with the fundamental difference that the value assigned to each \ac{PS} is randomly chosen. The beamforming matrix of \eqref{eq:beamformvect} becomes
\begin{equation}\label{eq:randbeamvect}
\mathbf{B}\left(f,\bm{\theta}_0 \right)=\mathbf{B}=\text{diag}\left(e^{j\,  {\upsilon_1}}, e^{j\,  {\upsilon_2}}, \ldots,\, e^{j\,  {\upsilon_i}}, \ldots,\, e^{j\,  {\upsilon_{\Ntx}}} \right)
\end{equation}
with  {$\upsilon_i \sim \mathcal{U}\left( 0, 2\pi \right)$}. Note that in this configuration the matrix in \eqref{eq:randbeamvect} does not depend on the frequency and on the steering direction, thus resulting in an array pattern with a random shape \cite{guerra2015position}.  In the simplest implementation,  {{random weighting}} could be realized using switches as discrete \acp{PS} randomly changing their status \cite{clemente2013wideband}. An important aspect is that, for both \ac{MIMO} and  {{random weighting}}, the rank of $\mathbf{B}$ is maximum and equal to $\Ntx$.

\subsection{Received Signal Model}
\label{sec:ch3_rxsignal}
In this section, a general framework for the received signal model is illustrated. The received signals are collected in a vector $\mathbf{r}(f) \!\!=\!\!\left[R_1(f),\, \hdots,\, R_m(f),\, \hdots,\, R_{\Nrx}(f) \right]^{\scriptscriptstyle \text{T}}$, where $\Rxm(f)=\mathcal{F}\left\{\rxm (t) \right\}$ is evaluated in $\Tob$ and $\rxm (t)$ is the equivalent low-pass received signal at the $m$th receiving antenna.  {Specifically, the received signal can be written as in \eqref{eq:rxsignal}. The receiving and transmitting direction matrices for the inter-antennas delays and {\ac{Tx}} orientation are given by
\begin{align}\label{eq:atar}
\! \!\! \!&\mathbf{a}^{\rx}(f, \aoal)=\left[e^{j\, \gamma_1^\rx},\,\hdots,\,e^{j\, \gamma_m^\rx},\,\hdots,\,e^{j\, \gamma_{\Nrx}^\rx} \right]^{\scriptscriptstyle \text{T}} \\
\! \!\! \!&\mathbf{A}^{\tx}(f,  \doal,\orien) \! =\! \text{diag} \left(e^{-j\, \gamma_1^\tx},\,\hdots,\,e^{-j\, \gamma_i^\tx},\,\hdots,\,e^{-j\, \gamma_{\Ntx}^\tx}\right) \label{eq:atar2}
\end{align}
where $\gamma_{i/m}^{\text{tr}}=2\, \pi\, \left( f+\fc\right)\, \tau_{i/m}^\text{tr} \left(\bm{\theta}_l^\text{tr}, \bm{\vartheta}^{\text{tr}} \right)$;} while  $\mathbf{c}(f,\tau_l) \!\!=\!\! c_l\, \mathbf{1}_{1 \times \Ntx}$ is the $1 \times \Ntx$ channel vector  whose generic element is $c_l\!\!=\!\!a_l\,e^{-j\, 2\, \pi \, (f+\fc) \, \taul}\!\!=\!\!{\alpha}_l\,e^{-j\, 2\, \pi \, f \, \taul} $. 
Specifically, the dominant \ac{LOS} component related to direct path (i.e., $l \!\!=\!\! 1$) is considered deterministic while, for $l > 1$, $ \alpha_{l} \sim \mathcal{CN} \left(0, \sigma_l^{ 2} \right) $ is a circularly symmetric Gaussian \ac{RV} statistically modelling the $l$th \ac{MPC} \cite{molisch2007wireless}. 
  Finally, $\mathbf{x}(f)\!\!=\!\!\left[{X}_1(f),\, \ldots,\, {X}_m(f),\, \ldots,\, {X}_{\Nrx}(f) \right]^{\scriptscriptstyle \text{T}}$ is the set of useful received signals and $\mathbf{n}(f)\!\!=\!\!\left[{N}_1(f),\, \ldots,\, {N}_m(f),\, \ldots,\, {N}_{\Nrx}(f) \right]^{\scriptscriptstyle \text{T}}$ is the noise vector with $\Noise(f)=\mathcal{F}\left\{\noise(t) \right\}$, with
 $\noise(t) \sim \mathcal{CN} \left(0, \No\right)$ being a circularly symmetric, zero-mean, complex Gaussian noise.
For further convenience, define $\nu_t= {\Etot}/\No=\nu \, \Ntx$, with  $\nu= E/\No$. The (total) \ac{SNR} at each receiving antenna element is  
$\mathsf{SNR}_{\text{t}}= \Ntx \mathsf{SNR}_1$, 
where $\mathsf{SNR}_1=\left(a_1\right)^2\,\nu$ represents the \ac{SNR} component related to the direct path between a generic couple of TX-RX antenna elements.

\section{Position and Orientation Error Bound}
\label{sec:posbound}
\subsection{Unknown Parameters}
The aim of the system is to estimate the position $\pt$ of the {\ac{Tx}}  and its orientation $\orien$ starting from the set of received waveforms $\mathbf{r}(f)$. 

In this context, \acp{MPC} and the residual {time synchronization} error represent nuisance parameters when evaluating the ultimate performance of the estimator. Thus, the unknown parameters vector is defined as
\begin{align}\label{eq:trueparametervector}
&\bm{\psi}=\left[\mathbf{q}^{\scriptscriptstyle \text{T}}, \,\bm{\kappa}^{\scriptscriptstyle \text{T}},\, \sincro \right]^{\scriptscriptstyle \text{T}} 
\end{align}
where  {the parameters of interest related to localization and orientation} are collected in $\mathbf{q}=\left[\left(\mathbf{p}^\tx\right)^{\scriptscriptstyle \text{T}}, \, \left(\bm{\vartheta}^\tx\right)^{\scriptscriptstyle \text{T}}\right]^{\scriptscriptstyle \text{T}}$, and the multipath parameters in  {$\bm{\kappa}=\left[\bm{\kappa}_1^{\scriptscriptstyle \text{T}},\bm{\kappa}_2^{\scriptscriptstyle \text{T}}, \ldots, \bm{\kappa}_l^{\scriptscriptstyle \text{T}}, \ldots, \bm{\kappa}_L^{\scriptscriptstyle \text{T}} \right]^{\scriptscriptstyle \text{T}}$}, with
\begin{align}\label{eq:multipathparam}
 \bm{\kappa}_l=\begin{cases}
\left[a_1 \right]^{\scriptscriptstyle \text{T}}\qquad\quad\,\,\, \text{if $l=1$} \\
\left[\alpharl, \,\, \alphail \right]^{\scriptscriptstyle \text{T}}\quad \text{if $l>1$}.
\end{cases}
\end{align}
The terms $\alpharl=\Re\left\{\alpha_l \right\}$ and $\alphail=\Im\left\{\alpha_l \right\}$ indicate the real and imaginary part of the complex channel coefficient, respectively \cite{godrich2009analysis,witrisal2012performance}. The {time synchronization} error is modeled as independent Gaussian zero-mean \ac{RV} with standard deviation $\sigma_{\epsilon}^2$.  Note that the nuisance parameters $\bm{\psi_\text{r}}=\left[\bm{\kappa}_2^{\scriptscriptstyle \text{T}},\, \ldots,\, \bm{\kappa}_L^{\scriptscriptstyle \text{T}} ,\, \sincro \right]^{\scriptscriptstyle \text{T}} $ are described statistically (\textit{a-priori} information available) whereas $\bm{\psi_\text{nr}}=\left[\mathbf{q}^{\scriptscriptstyle \text{T}}, \,a_1\right]^{\scriptscriptstyle \text{T}} $ are treated as deterministic (no \textit{a-priori} information available).

{In the following, we will discern among two different cases based on the orientation awareness. Specifically, we refer to the \textit{orientation-unaware} case for indicating the situation in which the {\ac{Tx}} orientation is not known at \ac{Rx}.  Contrarily, the \textit{orientation-aware} case is the opposite situation in which the orientation is exactly known at the {\ac{Rx}} side and it can be removed from the list of unknown parameters in \eqref{eq:trueparametervector}.}  {Moreover, we suppose that an initial search is conducted by the {\ac{Tx}} in order to coarsely estimate its own position and orientation with respect to the {\ac{Rx}}. Consequently, the beamforming weights can be communicate to the {\ac{Rx}} by exploiting the communication link dedicated to data exchange.} 
\subsection{\ac{CRB} General Expression}
The performance of any unbiased estimator $\widehat{\bm{\psi}}=\widehat{\bm{\psi}}\left(\mathbf{r}\left( f\right) \right)$ can be bounded by the hybrid \ac{CRB} defined as \cite{van2004detection}
\begin{equation}
\mathbb{E}_{\mathbf{r}, \bm{\psi}_\text{r}} \left\{\left[\widehat{\bm{\psi}}-\bm{\psi} \right] \left[\widehat{\bm{\psi}}-\bm{\psi}\right]^{\scriptscriptstyle \text{T}} \right\} \succeq \FIMt^{-1}=\mathsf{CRB}\left({\bm{\psi}} \right)
\end{equation}
where $\FIMt$ is the Bayesian \ac{FIM} defined as
\begin{align}\label{eq:FIM}
&\!\!\FIMt \triangleq - \mathbb{E}_{\mathbf{r}, \bm{\psi}_\text{r}} \left\{\nabla_{\bm{\psi}\bm{\psi}}^2 \, \loglik \right\}  =\FIMtd + \FIMtp \,  \nonumber \\
&\!\!\!\!=\!\left[\begin{array}{c:cc}
\mathbf{J}_{\mathbf{q}\mathbf{q}}^\text{d}&\mathbf{J}_{\mathbf{q}\bm{\kappa}}^\text{d}&\mathbf{J}_{\mathbf{q}\sincro}^\text{d} \\ \hdashline
\mathbf{J}_{\bm{\kappa} \mathbf{q}}^\text{d}&\mathbf{J}_{\bm{\kappa} \bm{\kappa}}^\text{d}+\mathbf{J}_{\bm{\kappa} \bm{\kappa}}^\text{p}&\mathbf{J}_{\bm{\kappa}\sincro}^\text{d} \\ 
\mathbf{J}_{\sincro \mathbf{q}}^\text{d}&\mathbf{J}_{\sincro \bm{\kappa}}^\text{d}&\mathbf{J}_{\sincro \sincro}^\text{d}+\mathbf{J}_{\sincro \sincro}^\text{p} 
\end{array}\right]\! \!=\!\!\left[\begin{array}{ll}
\mathbf{A} & \mathbf{C} \\ 
\mathbf{C}^{\scriptscriptstyle \text{H}} & \mathbf{D}
\end{array}\right] .
\end{align}
The symbol $\nabla_{\bm{\psi}\bm{\psi}}^2=\left( {\partial^2}/{\partial \bm{\psi} \partial \bm{\psi}} \right)$ denotes the second partial derivatives with respect to the elements in $\bm{\psi}$ and 
\begin{align}\label{eq:jtheta}
&\FIMtd =- \mathbb{E}_{\mathbf{r}, \bm{\psi}_\text{r}} \left\{\nabla_{\bm{\psi}\bm{\psi}}^2\, \loglikcond \right\} \quad  \nonumber \\
&\FIMtp=- \mathbb{E}_{\bm{\psi}_\text{r}} \left\{\nabla_{\bm{\psi}\bm{\psi}}^2\, \logprio \right\}
\end{align}
are the \ac{FIM} related to data and the \ac{FIM} containing the \textit{a-priori} statistical information on the parameters, respectively.

Since the observations at each receiving antenna element are independent, the log-likelihood function $\loglikcond$ can be written as 
\begin{equation}
\loglikcond  \propto -\frac{1}{\No} \sumrxm \int_{W} \lvert R_{m}(f)-X_m(f) \rvert^2 \, df.
\end{equation}
Moreover, based on the statistical information of $\bm{\psi}_\text{r}$, it is possible to derive the \textit{a-priori} probability density function of parameters $\bm{\psi}_\text{r}$ whose expression is reported in Appendix~\ref{appA}. All \ac{FIM} elements are reported in details in Appendixes A and B.

Finally, by using the Schur complement, the \ac{CRB} expression related to the localization and orientation estimation error can be easily derived as

\begin{equation}\label{eq:crb11}
\mathsf{CRB}\left(\mathbf{q} \right)=\left(\mathbf{A}- \mathbf{C}\,\,\mathbf{D}^{-1}\,\,\mathbf{C}^{\scriptscriptstyle \text{H}} \right)^{-1} \, .
\end{equation}

Equation \eqref{eq:crb11} is a general bound valid for different set-up (\ac{MIMO}, timed, phased and {random weighting} arrays) and accounting for signal weights quantization effects, {time synchronization} mismatch and  multipath. Specialized expressions can be derived from  \eqref{eq:crb11} for specific cases to get insights on the key parameters affecting the performance as will be done in {Sec.~\ref{sec:idealscenario}}.

{
\section{On the CRB tightness in massive array regime}
\label{sec:crbtight}
It is well known that the \ac{CRB} is a meaningful metric when the global ambiguities are negligible  \cite{van2004detection}. Such a condition is satisfied when operating at high \ac{SNR} (asymptotic \ac{SNR} regime) but, unfortunately, the required high \acp{SNR} cannot be in general obtained, especially at high frequencies. \\
\indent Therefore, in the following, we demonstrate that the global ambiguities can be made negligible without imposing the \ac{SNR} to be very large by letting the antenna array being massive (\textit{massive array regime}). In particular, we aim to show that, under random \ac{Rx} array orientations, the number of geometric configurations in which the ambiguities are not negligible vanishes as the number of receiving antennas increases.  \\
\indent To this purpose, the \ac{AF} is a powerful tool to investigate the presence of ambiguities especially used in radar systems and, it can be derived from the \ac{ML} discarding the thermal noise component \cite{san2007mimo_c}.\\
Let define the normalized \ac{AF} as
\begin{align}\label{eq:AFnorm}
 \mathsf{AF}\left(\mathbf{p},\tilde{\mathbf{p}} \right)& =\Bigg\lvert \, \frac{T_\text{obs}}{\Ntx\, \Nrx} \, \int_{W} \mathbf{x}^{\scriptscriptstyle \text{H}}(f,\mathbf{p}) \mathbf{x}(f,\tilde{\mathbf{p}} ) \, df\, \Bigg\rvert^2   
\end{align}
where $\mathbf{p}$ is the true \ac{Tx} position, $\tilde{\mathbf{p}}$ is a test position and, $\mathbf{x}$ is the useful signal vector reported in \eqref{eq:rxsignal}.
Asymptotically for $\Nrx \rightarrow \infty$ (\textit{massive array regime}), for the weak law of the large number \cite{van2000asymptotic_c},  we can write
\begin{align}\label{eq:AFeq}
 \mathsf{AF}\left(\mathbf{p},\tilde{\mathbf{p}} \right) \xrightarrow[]{\,P \,} \Bigg\lvert \, \frac{T_\text{obs}}{\Ntx\, \Nrx} \, \int_{W}  \mathbb{E}\left[ \,\mathbf{x}^{\scriptscriptstyle \text{H}}(f,\mathbf{p}) \mathbf{x}(f,\tilde{\mathbf{p}} )\right] \, df\, \Bigg\rvert^2 
\end{align}
where the operator $\xrightarrow[]{P}$ indicates the convergence in probability.
\\
In the following, we will consider the free-space and the multipath cases, separately, in order to show how the sidelobes level behaves in the massive array regime. The analysis  in non-massive array regime is considered in Sec. VIII.
\subsection{Free-space Scenario}
\label{sec:freespace_AF}
Here we focus our attention to the free-space scenario (i.e., $l=k=1$). In this case, the expectation term in \eqref{eq:AFeq} becomes
\begin{align}\label{eq:ris_par}
& \frac{1}{\Ntx\, \Nrx} \mathbb{E}\left[ \,  \mathbf{x}^{\scriptscriptstyle \text{H}}(f,\mathbf{p}) \mathbf{x}(f,\tilde{\mathbf{p}} ) \right] \propto \frac{1}{\Ntx \, \Nrx}\, \mathbb{E} \left[\mathbf{H}(\mathbf{p}, \tilde{\mathbf{p}}) \right]
\end{align}
where $\mathbf{H}(\mathbf{p}, \tilde{\mathbf{p}})$,  is a $\Ntx \times \Ntx$ matrix  whose generic element is given by
\begin{align}
\left[\mathbf{H}(\mathbf{p}, \tilde{\mathbf{p}})\right]_{i,j} &= \lvert a_1 \rvert^2 \, e^{-j\, 2\, \pi\, (f+\fc)\Delta \tau_1(\mathbf{p}, \tilde{\mathbf{p}})}\, \tilde{\omega}_i\, \tilde{\omega}_j^* \nonumber \\
&\times \, e^{j \Psi_{ij}^{(1,1)}(\mathbf{p}, \tilde{\mathbf{p}})}\, \sum_{m=1}^{\Nrx} \, \mathbb{E}\left[  e^{j \Psi_{m}^{(1,1)}(\mathbf{p}, \tilde{\mathbf{p}}, \bm{\vartheta}^\rx)} \right] \,   \nonumber \\
& = \begin{cases}
&\lvert a_1 \rvert^2 \Nrx \,  e^{j \Psi_{ij}^{(1,1)}(\mathbf{p}, \tilde{\mathbf{p}})}  \quad \text{$\mathbf{p}=\tilde{\mathbf{p}}$} \\
& 0 \qquad\qquad\qquad\qquad\quad  \text{otherwise}
\end{cases}
\end{align}
where we have defined $\Delta \tau_1(\mathbf{p}, \tilde{\mathbf{p}})= \tau_1(\mathbf{p})-\tau_1(\tilde{\mathbf{p}})$, $\Psi_{ij}^{(1,1)}(\mathbf{p}, \tilde{\mathbf{p}})= \gamma_i^\tx(\mathbf{p}, \bm{\vartheta}^\tx)-\gamma_j^\tx(\tilde{\mathbf{p}}, \bm{\vartheta}^\tx)$ and $\Psi_{m}^{(1,1)}(\mathbf{p}, \tilde{\mathbf{p}}, \bm{\vartheta}^\rx)= -\gamma_m^\rx(\mathbf{p}, \bm{\vartheta}^\rx)+\gamma_m^\rx(\tilde{\mathbf{p}}, \bm{\vartheta}^\rx) $ which depends on the \ac{Rx} array orientation. Note that $\Psi_{m}^{(1,1)}(\mathbf{p}, \tilde{\mathbf{p}}, \bm{\vartheta}^\rx)=0$ for $\mathbf{p}=\tilde{\mathbf{p}}$ regardless the \ac{Rx} orientation. On the other side, when $\mathbf{p}\neq \tilde{\mathbf{p}} $, in the presence of a large number of antenna elements ($\Nrx \rightarrow \infty$) and considering random \ac{Rx} orientations, the inter-antenna phase terms $\Psi_{m}(\mathbf{p}, \tilde{\mathbf{p}}, \bm{\vartheta}^\rx)$ can be modeled as independent \acp{RV} uniformly distributed in $\left[ 0,2\pi\right)$. In fact, different geometric configurations permit to span all the angles especially when large arrays are considered \footnote{{The goodness of the fitting with a uniform distribution has been validated through simulations.}}.
\\
\indent This means that the percentage of geometrical configurations of the \ac{Rx} for which the ambiguities are not negligible (i.e., $\mathsf{AF}(\mathbf{p}, \tilde{\mathbf{p}}) \rightarrow 0$ when $\mathbf{p} \neq \tilde{\mathbf{p}}$), vanishes as $\Nrx$ increases.
\\
\indent In other words, the conditions that permit to operate in the non-ambiguity region during the CRB evaluation are twofold: the first is to increase the SNR (high-SNR regime) by keeping the number of antennas fixed, whereas the second fixes the SNR (even not extremely large) and let the number of antennas explode.
\subsection{Multipath Scenario}
\label{sec:MP_AF}
This section aims at investigating if the \ac{CRB} still remains a meaningful metric in the presence of multipath. To this purpose, we consider the normalized \ac{AF} by putting in evidence the multipath contribution, as
\begin{align}\label{eq:afnormmultipath}
 \mathsf{AF}\left(\mathbf{p},\tilde{\mathbf{p}} \right)& =\Bigg\lvert \, \frac{T_\text{obs}}{\Ntx\, \Nrx} \, \int_{W} \mathbf{x}^{\scriptscriptstyle \text{H}}(f,\mathbf{p}) \mathbf{x}(f,\tilde{\mathbf{p}}) \, df\, \Bigg\rvert^2   \nonumber \\
&=\Bigg\lvert \, \frac{T_\text{obs}}{\Ntx\, \Nrx} \, \int_{W} \left( \mathbf{x}_1(f,\mathbf{p})+\mathbf{x}_{l>1}(f,\mathbf{p}) \right)^{\scriptscriptstyle \text{H}} \nonumber \\
& \times  \left( \mathbf{x}_1(f,\tilde{\mathbf{p}})+\mathbf{x}_{l>1}(f,\tilde{\mathbf{p}}) \right)  \, df\, \Bigg\rvert^2   \nonumber \\
&=\Bigg\lvert \, \int_W  \frac{f_\text{AWGN}\left(\mathbf{p},\tilde{\mathbf{p}} \right)}{\Ntx\, \Nrx}  +   \frac{f_\text{MP}\left(\mathbf{p},\tilde{\mathbf{p}} \right)}{\Ntx \Nrx}  df  \, \Bigg\rvert^2 \, 
\end{align}
where $\mathbf{x}_1(f,\mathbf{p})$ and $\mathbf{x}_{l>1}(f,{\mathbf{p}})$ indicate the expected received (noise-free) signal due to the direct path and multipath, respectively.
Given the expression in \eqref{eq:afnormmultipath}, the following asymptotic analysis aims at verifying that the number of times the multipath impacts on the \ac{AF} shape is negligible compared to the number of times it has not an effect at all, provided that the number of {\ac{Rx}} antennas goes to infinity and that random array orientations are considered. More precisely, recalling the weak law of the large numbers, it is
\begin{equation}\label{eq:problem}
\frac{f_\text{MP}\left(\mathbf{p},\tilde{\mathbf{p}} \right)}{\Ntx \Nrx}  \xrightarrow[]{\,\,\, P \,\,\, } \frac{1}{\Ntx\, \Nrx} \mathbb{E}\left[f_\text{MP}\left(\mathbf{p},\tilde{\mathbf{p}} \right) \right]\, 
\end{equation}
where we aim at verifying that the right-hand side of \eqref{eq:problem} is $0$ for $\mathbf{p} \neq \tilde{\mathbf{p}}$, meaning that \ac{AF} sidelobes depending on multipath disappear when $\Nrx$ is large and random orientations are considered.\\  
The expectation argument in \eqref{eq:problem} is given by
\begin{align}\label{eq:mpexp}
\mathbb{E}\left[f_\text{MP}\left(\mathbf{p},\tilde{\mathbf{p}} \right) \right] &=  \mathbb{E} \left[\mathbf{x}_{l>1}^\text{H} (f,\mathbf{p})  \mathbf{x}_1(f,\tilde{\mathbf{p}})\right] \nonumber \\
&+ \mathbb{E}\left[\mathbf{x}_1^\text{H}(f,\mathbf{p})  \mathbf{x}_{l>1} (f,\tilde{\mathbf{p}})\right] \nonumber \\
& + \mathbb{E}\left[\mathbf{x}_{l>1}^\text{H}(f,\mathbf{p})  \mathbf{x}_{l>1} (f,\tilde{\mathbf{p}})\right].
\end{align}
Treating separately the terms in \eqref{eq:mpexp}, we have 
\begin{align}
&\mathbb{E}\left[ \mathbf{x}_1^\text{H}(f,\mathbf{p})  \mathbf{x}_{l>1} (f,\tilde{\mathbf{p}})    \right]= \nonumber \\
&= \sum_{mij} \sum_{k=2}^{L} \alpha_1^* \,\alpha_k \, S_i(f)S_j^*(f)\, \tilde{\omega}_i\, \tilde{\omega}_j^*\,e^{-j  2\, \pi\, f \, \tau_k}\, e^{-j \,\Psi_{ij}^{(1,k)}\left( \mathbf{p}, \tilde{\mathbf{p}} \right)}\nonumber \\
& \times  \mathbb{E}\left[ e^{j \Psi_m^{(1,k)}(\mathbf{p}, \tilde{\mathbf{p}})} \right] = 0 \qquad \forall \tilde{\mathbf{p}}
\end{align}
where $\sum_{mij}=\sum_{m=1}^{\Nrx}\sum_{i=1}^{\Ntx}\sum_{j=1}^{\Ntx}$, $\Psi_m^{(1,k)}(\mathbf{p}, \tilde{\mathbf{p}})=-\gamma_m(\bm{\theta}_1, \bm{\vartheta}^\rx) + \gamma_m(\bm{\theta}_k, \bm{\vartheta}^\rx) $, $\Psi_{ij}^{(1,k)}=-\gamma_i(\bm{\theta}_1(\mathbf{p}))+\gamma_j(\bm{\theta}_1(\tilde{\mathbf{p}}))$, and
$\mathbb{E}\left[ e^{-j \left( 2\, \pi\, f \, \tau_k+\, \Psi_m^{(1,k)}(\mathbf{p}, \tilde{\mathbf{p}})\right)} \right] = 0 $ as the phases are assumed uniformly distributed between $0$ and $2 \, \pi$. Similar considerations are valid for $\mathbb{E} \left[\mathbf{x}_{l>1}^\text{H} (f,\mathbf{p})  \mathbf{x}_1(f,\tilde{\mathbf{p}})\right]$. 
\\
Finally, consider the last term in \eqref{eq:mpexp}, i.e.
\begin{align}\label{eq:mp_terms}
&\mathbb{E}\left[ \mathbf{x}_{l>1}^\text{H}(f,\mathbf{p})  \mathbf{x}_{l>1} (f,\tilde{\mathbf{p}})    \right]= \nonumber \\
&= \sum_{mij}\sum_{l=2}^{L}\sum_{k=2}^{L}\, S_i(f)\, S_j(f)\,\tilde{\omega}_i\, \tilde{\omega}_j^*\, e^{-j\, \Psi_{ij}^{(l,k)} \left( \mathbf{p}, \tilde{\mathbf{p}} \right)} \nonumber \\
&\times \alpha_l\, \alpha_k^*\,e^{-j\, 2\, \pi\, f \, \Delta \tau_{lk}} \, \mathbb{E}\left[e^{-j\, \Psi_m^{(l,k)} \left( \mathbf{p}, \tilde{\mathbf{p}} \right)} 	\right]
\end{align}
where $\Delta \tau_{lk}=\tau_l-\tau_k$, $\Psi_m^{(l,k)} \left( \mathbf{p}, \tilde{\mathbf{p}} \right)=\gamma_m^\rx(\bm{\theta}_l, \mathbf{p}, \bm{\vartheta}^\rx)- \gamma_m^\rx(\bm{\theta}_k, \tilde{\mathbf{p}}, \bm{\vartheta}^\rx)$, $\Psi_{ij}^{(l,k)} \left( \mathbf{p}, \tilde{\mathbf{p}} \right)=\gamma_i^\tx(\bm{\theta}_l, \mathbf{p})- \gamma_i^\tx(\bm{\theta}_k, \tilde{\mathbf{p}})$.\\
In this case, since it holds 
\begin{align}\label{eq:effect}
& \mathbb{E}\left[  \,e^{-j\, \Psi_m^{(l,k)} \left( \mathbf{p}, \tilde{\mathbf{p}} \right)}\right]  \nonumber \\
& =\begin{cases}
&0 \quad \text{if $l \neq k, \quad \forall \tilde{\mathbf{p}}$} \\
&1  \quad \text{if $l = k, \quad  \mathbf{p} = \tilde{\mathbf{p}}  $ }\\
&\mathbb{E}\left[ e^{-j\, \Psi_m^{(l,l)} \left( \mathbf{p}, \tilde{\mathbf{p}} \right)} 	\right] =0 \quad \text{if $l = k, \quad  \mathbf{p} \neq \tilde{\mathbf{p}} $\, , }
\end{cases}
\end{align}
it follows that \eqref{eq:mp_terms} is equal to $0$ for $\mathbf{p} \neq \tilde{\mathbf{p}} $, i.e. in all those cases in which a global ambiguity can arise.\\
\indent The obtained result shows that the global ambiguities due to the multipath are, on average, negligible. Nevertheless, the effect of multipath still remains in the correspondence of the true peak of the \ac{AF}, i.e., that for $\mathbf{p}=\tilde{\mathbf{p}}$, as reported in \eqref{eq:effect} for $l=k$. Consequently, even if we can state that the \ac{CRB} is a valid metric in establishing the ultimate performance provided that $\Nrx$ is sufficiently large, the effect of multipath on the localization accuracy necessitates to be investigated. Specifically, Sec.~\ref{sec:mp_loc_ac} analyzes the effect of multipath from a localization accuracy point-of-view.
}
%
\section{{ Free-Space Localization Bound }}
\label{sec:idealscenario}
Here we provide an example on how the general expression \eqref{eq:crb11} can be simplified in absence of beamforming weights errors and \acp{MPC}. Specifically, in free-space conditions, \eqref{eq:FIM} can be reduced to 
\begin{equation} \label{eq:FIMetaThetaAWGN}
\FIMtd=\FIMt=\left[\begin{array}{cc} 
\mathbf{J}_{\mathbf{q}\mathbf{q}}& \mathbf{J}_{\mathbf{q}a_1} \\ 
\mathbf{J}_{a_1 \mathbf{q}}& {J}_{a_1 a_1} \,
\end{array}\right]=\left[\begin{array}{cc} 
\mathbf{J}_{\mathbf{q}\mathbf{q}}^{\text{ {FS}}} &  {\mathbf{0}} \\ 
 {\mathbf{0}}  & {J}_{a_1 a_1}\,
\end{array}\right]
\end{equation}
where its elements are reported in Appendix~\ref{appB} and where  the superscript $^\text{d}$ is omitted as in this case all the parameters to estimate are deterministic.   {For readability convenience, we report here the expression of the \ac{FIM} related to the localization parameters, that is:
\begin{align}\label{eq:Jqq_main}
J_{q_b\, q_a}=& 8\, \pi^2\, \nu \, a_1^2 \, \sum_{mij} \Re \left\{\tilde{b}_{ij}^\text{c}\,\xi_{ij}^{(1,1)}\,  \chi_{ij}^{(1,1)}(2)\,  \right\} \nonumber \\
&\times \nabla_{q_a}\left( \tau_{im1}\right)  \nabla_{q_b}\left( \tau_{jm1} \right)
\end{align}
where  {$q_{a/b}$ are two elements in the set $\left\{x,\, y,\, z,\, \orient ,\, \orienp \right\}$, and}
\begin{align}\label{eq:chiij2}
&\chi_{ij}^{(1,1)}(2)\!\!=\!\!\int_W\!\!\!\tilde{b}_{ij}(f)\left(f+\fc \right)^2\! e^{-j\, 2\, \pi\, f\, \Delta \tau_{ij}^{(1,1)}}P_i(f)\, P_j^*(f) \,df  
\end{align}
with $\Delta \tau_{ij}^{(1,1)}=\tau_{im1}-\tau_{jm1}$, $\xi_{ij}^{(1,1)}=e^{-j\, 2\, \pi\, \fc\, \Delta \tau_{ij}^{(1,1)}}$, $\tilde{b}_{ij}(f)=\tilde{b}_i(f)\, \tilde{b}_j^*(f)$, and $\tilde{b}_{ij}^\text{c}=\tilde{b}_i^\text{c}\, \left(\tilde{b}_j^\text{c}\right)^*$. In \eqref{eq:Jqq_main}, the derivatives translate the \ac{TOA} and \ac{DOA} in position and orientation information. In particular, for the position we have 
\begin{align}\label{eq:der_toa_q0}
& \nabla_p\left(\tau_{im1} \right)= \frac{1}{c}\, \left\{ c\,\nabla_p\left( \tau_1\right)+\nabla_p\left( \bm{\theta}_1\right) \left[\mathbf{p}_i^\tx \left(\bm{\vartheta}^\tx\right)-\mathbf{p}_m^\rx\left(\bm{\vartheta}^\rx \right) \right] \right\}.
\end{align}
The term $\nabla_p\left(\tau_1 \right)$ expresses the dependence of the position from the direct path \ac{TOA}; while 
\begin{align}\label{eq:etaq1}
\nabla_p\left(\bm{\theta}_1 \right)=& \nabla_p\left({\theta}_1 \right)\, \cos(\theta_1) \left[\begin{array}{c}
\cos(\phi_1) \\ \sin(\phi_1) \\ - \tan(\phi_1)
\end{array}\right]^{\scriptscriptstyle \text{T}} \nonumber \\
&+ \nabla_p\left({\phi}_1 \right)\, \sin(\theta_1) \left[\begin{array}{c}
-\sin(\phi_1) \\ \cos(\phi_1) \\ 0
\end{array}\right]^{\scriptscriptstyle \text{T}}
\end{align}
includes the dependence of the position from the \ac{DOA} information. Finally, for what the orientation information is regarded, we have
\begin{align}\label{eq:nablaorie}
&  \nabla_{\bm{\vartheta}^\tx}(  {\tau_{im1}})= \nabla_{\bm{\vartheta}^\tx}(  {\tau_{i}^\tx\left(\bm{\theta}_1, \bm{\vartheta}^\tx \right)})= \frac{1}{c}\, \mathbf{d}\left( \bm{\theta}_1 \right)\, \nabla_{\bm{\vartheta}^\tx}\left(\mathbf{p}_i^\tx\left( \bm{\vartheta}^\tx\right) \right).
\end{align}
By further analyzing \eqref{eq:Jqq_main}, one can notice the dependence of the \ac{FIM} from the beamforming weights given by the coefficients $\tilde{b}_{ij}^\text{c}$ and $\tilde{b}_{ij}(f)$.\\
Given the \ac{FIM} in \eqref{eq:Jqq_main} and starting from \eqref{eq:FIMetaThetaAWGN}, it can be easily found that for beamforming and \ac{MIMO} it is
}
\begin{equation}\label{eq:crbawgn}
\mathsf{CRB}^\text{FS}\left(\mathbf{q} \right)=\!\left(\mathbf{J}_{\mathbf{q}\mathbf{q}}^\text{FS}\right)^{-1}= \left(\breve{\mathbf{J}}_{\mathbf{q}\mathbf{q}}^\text{FS}\,\, \mathbf{G} \right)^{-1}
\end{equation}
where  {we have separated} the effect of signal design $\breve{\mathbf{J}}_{\mathbf{q}\mathbf{q}}^\text{FS}$, i.e., that related to \eqref{eq:chiij2}, from that of the geometry $\mathbf{G}$, i.e., that related to \eqref{eq:der_toa_q0}-\eqref{eq:nablaorie}. Specifically for timed arrays,  we have
\begin{align}\label{eq:crbawgn_elem}
&\breve{\mathbf{J}}_{\mathbf{q}\mathbf{q}}^\text{FS}={8 \pi^2 \,\mathsf{SNR}_1} \left({\beta^2}+\fc^2 \right),\,\,\mathbf{G}=\sum_{mij}  \nabla_{\mathbf{q}\mathbf{q}}\left(\tau_{im1} ,\tau_{jm1} \right)
\end{align}
where  {$\nabla_{\mathbf{q}\mathbf{q}}\left(\tau_{im1} ,\tau_{jm1} \right)$ is a $5 \times 5$ matrix whose entries are given by $\nabla_{{q}_a}\left(\tau_{im1}\right) \nabla_{{q}_b}\left(\tau_{jm1} \right)$}, and $\beta$ is the baseband effective bandwidth of $p(t)$, defined as
\begin{align}
&\beta=\left(\int_{W}\, f^2\, \lvert P(f) \rvert^2\,df \right)^{\frac{1}{2}}\,.
\end{align}

\noindent Similarly, for \ac{MIMO} arrays, it is possible to find
\begin{align}
&\breve{\mathbf{J}}_{\mathbf{q}\mathbf{q}}^\text{FS}= {8 \pi^2 \,\mathsf{SNR}_1} \left(\beta^2_i+\fc^2 \right), \,\, \mathbf{G}=\sum_{mi} \nabla_{\mathbf{q}\mathbf{q}}\left(\tau_{im1} ,\tau_{im1} \right) 
\end{align}
where $\sum_{mi}=\sum_{m=1}^{\Nrx}\sum_{i=1}^{\Ntx}$ and  $\beta^2_i=\frac{\beta^2}{\Ntx}$ is the squared baseband effective bandwidth of $\pimp(t)$. 

The matrix $\mathbf{G}$ provides, through derivatives, the relationship between the \ac{TOA} at each TX-RX antenna element couple and the {\ac{Tx}} position and orientation.  

To improve the comprehension of \eqref{eq:crbawgn}-\eqref{eq:crbawgn_elem}, in the next sections two particular cases of planar \ac{MIMO} and timed arrays will be discussed considering a fixed {\ac{Tx} and \ac{Rx}} orientation {i.e., $\orien=\bm{\vartheta}^\rx=\left[0, 0 \right]^{\scriptscriptstyle \text{T}}$}.  Note that the overall \ac{CRB} analysis is still valid for any orientation. In Secs.~\ref{sec:MIMOplanarArray} and \ref{sec:TIMEDplanarArray}, we choose a specific case just to provide some insights on how the number of transmitting and receiving antennas can impact the performance.  {In Appendix~\ref{appC}, the matrix $\mathbf{G}$ is evaluated, considering this specific array geometry.}

\subsubsection{ {Special Case: Planar MIMO Array}} 
\label{sec:MIMOplanarArray}
For the planar geometric configuration and in the \textit{orientation-unaware} case, the diagonal elements in the position and orientation \ac{CRB} matrix derived starting from \eqref{eq:crbawgn}-\eqref{eq:crbawgn_elem} and from \eqref{eq:Gresults1}, are given by 
\begin{align}\label{eq:crbMIMOpeo}
&\mathsf{CRB}\left(x \right)=\mathsf{CRB}\left(z \right)=\mathsf{CRB}_0\,\frac{12}{S\, \left(\Nrx-1\right)} \nonumber \\ 
&\mathsf{CRB}\left(y \right)=\frac{\mathsf{CRB}_0}{\Nrx} \nonumber \\
&\mathsf{CRB}\left(\orient \right)\!=\!\mathsf{CRB}\left(\orienp \right) \!=\!\mathsf{CRB}_0\,\frac{12\,\left(\Ntx+\Nrx-2\right)}{A_\text{rx} \left(\Ntx-1\right)\! \left(\Nrx-1\right)}
\end{align}
where $\mathsf{CRB}_0={c^2}/{\left(8 \pi^2 \, \mathsf{SNR}_{\text{t}}\, \left(\beta^2_i+\fc^2\right)\right)}$  is the \ac{CRB} of the ranging error one would obtain using single antenna, and $S={A^\rx}/{y^2}$ represents the ratio between the {\ac{Rx}} array area and the squared {\ac{Tx}-\ac{Rx}} distance. Note that  $\mathsf{CRB}_0$ depends on the carrier frequency $\fc$, on the shape of the pulse through $\beta_i^2$, on the received \ac{SNR}, and it does not depend on the number of transmitting antennas. The analytical derivation is reported in Appendix~\ref{appC}.  From \eqref{eq:crbMIMOpeo}, it is possible to remark that the \ac{CRB} of the estimation error in the $y$-coordinate is inversely proportional to the number of the receiving antenna elements accounting for the number of independent measurements available at the {\ac{Rx}}. Regarding the other two coordinates, a key parameter on the estimation accuracy is $S$ which is related to the ratio between the dimension of the {\ac{Rx}} array and the distance between the arrays: as this ratio becomes smaller (i.e., as the distance between the arrays becomes larger with respect to the array size), the positioning accuracy degrades. 
From \eqref{eq:crbMIMOpeo} it is also possible to notice that the accuracy in estimating the orientation depends both on the transmitting and receiving antennas. Specifically both $\Ntx$ and $\Nrx$ must be greater than one to make the orientation possible, whereas for the positioning, the constraint is only on the number of receiving elements that must be larger than $1$. Moreover, non-zero off-diagonal elements remark a correlation between the error on the estimation of  position and orientation. Specifically we have
\begin{align}
\mathsf{CRB}\left(z, \orient \right)&=\mathsf{CRB}\left(\orient, z \right)=\mathsf{CRB}\left(x, \orienp \right) =\mathsf{CRB}\left(\orienp, x  \right) \nonumber \\
&=\mathsf{CRB}_0 \frac{12}{S\, y\, \left(1-\Nrx\right)}.
\end{align}
Contrarily in the \textit{orientation aware} case, it can be found
\begin{align}\label{eq:CRBmimoknown}
&\mathsf{CRB}\left(x  \right)=\mathsf{CRB}\left(z\right)= \mathsf{CRB}_0\,\frac{12}{S\, \left(\Ntx+\Nrx-2\right)} \nonumber \\
& \mathsf{CRB}\left(y\right)=\frac{\mathsf{CRB}_0}{\Nrx}.
\end{align}
Note that when passing from a condition of \textit{orientation-unawareness} to that of \textit{orientation-awareness} the positioning accuracy increases, thanks to the additional information provided. In fact, the \ac{CRB} on $x$ and $z$ coordinates now depends also on the number of transmitting antennas.
\subsubsection{ {Special Case: Planar Timed Array}}
\label{sec:TIMEDplanarArray}

Differently from \ac{MIMO}, here in the \textit{orientation-unaware} case, the equivalent \ac{FIM} for position and orientation is singular meaning that it is not possible to jointly localize and determine the orientation using beamforming strategies. {Nevertheless, when multiple beams are generated \cite{alkhateeb2014channel_c,garcia2016location_c}, such singularity can be solved thus allowing the localization process, but at the prize of an increased scanning time if the beams are sequentially generated in time, or of a decreased \ac{SNR} if such beams are simultaneously formed. The investigation of this trade-off is out-of-the-scope of this paper.}
 
If  the {\ac{Tx}} orientation is a known parameter (\textit{orientation aware} case) and it is discarded from the estimation parameters vector, the elements of the position \ac{CRB} matrix result from  \eqref{eq:Gresults2}
\begin{align}\label{eq:crbP}
&\mathsf{CRB}\left(x \right)=\mathsf{CRB}\left(z \right)=\mathsf{CRB}_0\,\frac{12}{S}\frac{1}{\Ntx\,(\Nrx-1)} 
\nonumber \\
& \mathsf{CRB}\left(y \right)=\frac{\mathsf{CRB}_0}{\Ntx\,\Nrx}.
\end{align}
From \eqref{eq:crbP} it is possible to remark that the \ac{CRB} of the estimation error in the $y$-coordinate  is inversely proportional to $\Ntx$ and $\Nrx$: in fact, the $\Ntx$ term accounts for the \ac{SNR} enhancement due to the beamforming process while the $\Nrx$ term accounts for the number of independent measurements available at the {\ac{Rx}} (receiver diversity).  Note that when $\Nrx=1$, the localization along the $x$ and $z$ axes is not possible (only ranging in the $y$ direction), as for \ac{MIMO}. Refer to Appendix~\ref{appC} for more details related to the derivation of \eqref{eq:crbP}.
{
\section{Multipath effect on localization accuracy}
\label{sec:mp_loc_ac}
%
Once verified that the \ac{CRB} is a meaningful metric in different propagation conditions in Sec.~\ref{sec:MP_AF}, we now investigate the impact of \acp{MPC} on the localization performance for the considered scenario. In \cite{shen2010fundamental}, it is demonstrated that only the information related to the \textit{first-contiguous cluster}, i.e. the set of \acp{MPC} overlapped to the first path,  is relevant from a localization perspective in the asymptotic SNR regime. Here we show that under the asymptotic massive antenna regime, all the MPCs can be made negligible, included those belonging to the first-contiguous cluster. \\
The FIM in presence of multipath can be written as follows
\begin{equation}\label{eqFIMe2}
\FIMt=\left[\begin{array}{cc}
\mathbf{J}_{\mathbf{q}\mathbf{q}} & \mathbf{J}_{\mathbf{q}\bm{\kappa}}  \\ 
\mathbf{J}_{\bm{\kappa}\mathbf{q}}& {\mathbf{J}}_{\bm{\kappa}\bm{\kappa}}
\end{array}\right]
\end{equation}
where ${\mathbf{J}}_{\bm{\kappa}\bm{\kappa}}$ contains also the \textit{a-priori} information on \acp{MPC} statistics reported in Appendix A. Consequently, the \ac{CRB} for the multipath scenario can be formulated as
\begin{equation}\label{eq:CRB_MP}
\mathsf{CRB}\left(\mathbf{q} \right)=\left({\mathbf{J}}_{\mathbf{q}\mathbf{q}}-\mathbf{J}_{\mathbf{q}\bm{\kappa}}\, {\mathbf{J}}_{\bm{\kappa}\bm{\kappa}}^{-1} \, \mathbf{J}_{\bm{\kappa}\mathbf{q}} \right)^{-1}
\end{equation}
where all multipath information is gathered in $\mathbf{J}_{\mathbf{q}\bm{\kappa}}\, {\mathbf{J}}_{\bm{\kappa}\bm{\kappa}}^{-1} \, \mathbf{J}_{\bm{\kappa}\mathbf{q}}$. \\
Considering the average over different geometric configurations (e.g., the average over different \ac{Rx} orientations) and for large values of $\Nrx$, it is possible to show that the number of configurations where the multipath impacts the localization performance  compared to the number of configurations in which it does not influence the accuracy is negligible regardless the array architecture chosen. \\
Considering \eqref{eq:CRB_MP}, for the weak law of the large number (i.e., for $\Nrx \rightarrow \infty$), it holds 
\begin{equation}\label{eq:lim_prob}
\frac{1}{\Nrx \Ntx} \mathbf{J}_{\mathbf{q} \bm{\kappa}}  \xrightarrow[]{\,\,\, P \,\,\, }  \frac{1}{\Nrx \Ntx} \, \mathbb{E} \left[\mathbf{J}_{\mathbf{q} \bm{\kappa}}\right]\, 
\end{equation}
where we aim at demonstrating that
\begin{equation}\label{eq:lim_prob2}
\frac{1}{\Nrx \Ntx} \, \mathbb{E} \left[\mathbf{J}_{\mathbf{q} \bm{\kappa}}\right]=0 \,.
\end{equation}
In the presence of a large number of antenna elements and considering random \ac{Rx} orientations, the inter-antenna phase terms can be modeled as \acp{RV} uniformly distributed in $\left[ 0,2\pi\right)$. Under this assumption, we have
\begin{align}\label{eq:Jqkappa_exp}
&\mathbb{E}\left[ J_{q\,a_1 }  \right]= J_{q\,a_1}=0 \nonumber \\
&\mathbb{E}\left[ J_{ q \, \alpha_k^\Re} \right] = -4\, \pi\, a_1 \,\nu  {\sum_{mij}} \Im \left\{\tilde{b}_{ij}^\text{c}\,\mathbb{E}\left[ \,  \xi_{ij}^{(k,1)} \chi_{ij}^{(k,1)}(1)\right] \right\}  \nonumber \\
&\qquad\qquad \times \nabla_{q}\left( \tau_{jm1}\right) =0 \nonumber \\
&\mathbb{E}\left[ J_{ q \, \alpha_k^\Im} \right]= 4\, \pi\, a_1 \,\nu {\sum_{mij}} \Re \left\{\tilde{b}_{ij}^\text{c}\,\mathbb{E}\left[ \,  \xi_{ij}^{(k,1)} \chi_{ij}^{(k,1)}(1)\right] \right\}  \nonumber\\
&\qquad\qquad  \times \nabla_{q}\left( \tau_{im1}\right)   =0
\end{align}
where
\begin{align}\label{eq:exp_ext}
&\mathbb{E}\left[\xi_{ij}^{(k,1)} \chi_{ij}^{(k,1)}(1)\right] \propto  \mathbb{E}[ e^{-j\, 2\, \pi\, (f+\fc)\,(\Delta \tau_m^\rx(\bm{\theta}_1, \bm{\theta}_k))}]\, =0
\end{align}
with $\Delta \tau_m^\rx(\bm{\theta}_1, \bm{\theta}_k)=\tau_m^\rx(\bm{\theta}_1, \bm{\vartheta}^\rx)-\tau_{m}^\rx(\bm{\theta}_k, \bm{\vartheta}^\rx)$. Following similar considerations, it is straightforward to prove that the expectation of the $\mathbf{J}_{\bm{\kappa} \mathbf{q}}$ elements is zero. \\
The result in \eqref{eq:lim_prob} leads to the important conclusion that letting the antennas array be massive, i.e., large $\Nrx$, makes the set of geometric configurations significantly impacted by \acp{MPC} negligible, and the performance converges to that of the free space case. As a consequence, the CRB converges to the CRB averaged over the RX orientations for massive antenna arrays.
}

\section{Numerical Results}
\label{sec:numerical}

In this section, numerical results are reported considering different array schemes, multipath conditions and system non-idealities. Four array structures are analyzed: timed arrays  equipped with \acp{TDL} and \acp{PS}, phased and {random weighting} arrays using only \acp{PS}, and finally the \ac{MIMO} array in which orthogonal waveforms are transmitted and neither \acp{PS} nor \acp{TDL} are present. For what the antennas spatial deployment is regarded, planar arrays are considered as they represent the most conventional structure to be integrated in \acp{AP} and mobiles.  {Differently from Secs.~\ref{sec:MIMOplanarArray} and ~\ref{sec:TIMEDplanarArray}, here we consider the results averaged over the {\ac{Rx}} orientations if not otherwise indicated, and thus it will be possible to appreciate the impact of the array rotational angle on the localization performance}. {In the following figures, we indicate with Q the presence of quantization errors, with S the presence of a residual {time synchronization} error.} {
Moreover, we designate with:
\begin{itemize}
\item \textit{Fixed orientation}: the array configuration with the {\ac{Tx}} and the {\ac{Rx}} parallel to each other (i.e., $\orien=\bm{\vartheta}^\rx=\left[0,\, 0 \right]^{\scriptscriptstyle T}$), as described in Sec.~\ref{sec:idealscenario};
\item \textit{Averaged orientation}: the geometric configuration in which, for each Monte Carlo iteration, a different 3D {\ac{Rx}} array orientation is generated, and the \ac{CRB} is computed as the average over all the partial \ac{CRB} results computed at each cycle.
\end{itemize}
Finally, we recall that:
\begin{itemize}
\item \textit{Orientation-aware} indicates the case in which the {\ac{Tx}} orientation is known at the {\ac{Rx}} side and, thus, it is not considered in the parameters vector to be estimated;
\item \textit{Orientation-unaware}: the case in which the {\ac{Tx}} orientation is unknown at {\ac{Rx}} side and it has to be estimated together with the position. In the next figures, we will denote with O,  the \textit{orientation-unawareness} case.
\end{itemize}
}
\subsection{System Configuration}
\label{sec:ch3_sysconfig}

We consider a scenario with a single \ac{AP} equipped with a massive array, with the centroid placed in $\pr=[0,0,0]^{\scriptscriptstyle \text{T}}$, and a transmitting array antenna whose centroid is located in $\pt=[0,5,0]^{\scriptscriptstyle \text{T}}$ ($d = 5\,$m). 

As in the mathematical model, the {\ac{Rx}} has a perfect knowledge of the {\ac{Tx}} steering direction, and  the results are obtained for $\fc \!\!=\!\!60\,$GHz and $W\!\!=\!\!1\,$GHz (the signal duration is $\tp=1.6\,$ns) in free-space and multipath conditions. \ac{RRC} transmitted pulses centered at frequency $\fc=60\,$GHz and roll-off factor {of $0.6$} are adopted, being compliant with \ac{FCC} mask at $60$ GHz \cite{FCC60r}. A receiver noise figure of $N_\text{F}=4\,$dB and a fixed transmitted power of $P_\text{t}=10\,$mW  {are considered}, if not otherwise indicated.

The performance is evaluated in terms of  \ac{PEB} and  \ac{OEB} averaged over $N_\text{cycle}=500$ Monte Carlo iterations.
For each cycle, a different {3D} {\ac{Rx}} array orientation  {, i.e., $\bm{\vartheta}^\rx=\left[{\vartheta}^\rx,\,{\varphi}^\rx \right]^{\scriptscriptstyle \text{T}}$ ,} and multipath scenario are generated. Specifically, the receiving (transmitting) antennas are spaced apart of $d_\text{ant}=\lambda_L /2$, where $\lambda_{L}={c}/{f_L}$ and $f_L=\fc-W/2$.
When present, the \acp{PS} quantization errors are $\deltait \sim \mathcal{U} \left(-\pi/4, \pi/4\right)$ while the \acp{TDL} errors are $\Delta \tauit\sim \mathcal{U}  \left(0, d_\text{ant}/c \right)$. The standard deviation of the {time synchronization} error is set to $\sigma_{\epsilon}=1\,\text{ns}$.

When operating at \ac{mm-wave} frequencies, the path arrival time distributions can be described by a Poisson process and the inter-arrival times by an  exponential probability density function \cite{gustafson2014mm}. The paths arrival rate is set to $4\,$[1/ns] while the paths azimuth and elevation \ac{AOA} are modeled as uniformly distributed between $(0, 2\pi]$ and $(0, \pi]$, respectively. {Note that these values are also in line with those found in \cite{guerra2016delay_c,guidi2017eucap_c} where a \ac{mm-wave} measurements campaign using massive arrays for radar-based mapping purposes have been described.}

{
Before analyzing the \ac{MIMO} and beamforming localization performance, it is necessary to ensure that the comparison based on \ac{CRB} can be considered fair in terms of \ac{SNR} working regimes when operating with non-massive arrays. To this purpose, a threshold in terms of \ac{SNR} is derived in order to understand if ambiguities are significant or not and, hence, whether the \ac{CRB} can still be used as a performance metric for comparison. In this perspective, we still consider the \ac{AF} as a tool to investigate the performance of \ac{MLE} as a function of the probability of ambiguity, i.e., secondary lobes higher than the main lobe. In fact, the \ac{AF} main lobe determines the \ac{MSE} behaviour in the high \ac{SNR} region that is described by the \ac{CRB}, while the \ac{AF} sidelobes might generate ambiguities in case of large noise in the low \ac{SNR} region and that are not taken into account by the \ac{CRB}. The details of the aforementioned analysis are reported in Appendix D.\\
In our numerical results, for each tested configuration we verified that the \ac{SNR} level is above a threshold calculated to guarantee that the probability of ambiguity is less than $10^{-2}$.
}
\subsection{Results}
\label{sec:ch3_results}
The results of this section have been obtained as a function of $\Ntx$ and $\Nrx$; the array structure (i.e. timed, phased, random, \ac{MIMO}); the presence and absence of array beamforming and of a residual {time synchronization} error; and the multipath overlapping effect.

\subsubsection{Free space scenario}
\label{sec:ch3_awgnresults}
\paragraph{Results as a function of $\Ntx$}
\label{sec:Ntxresults}
\begin{figure}[t!]
\psfrag{PEB}[c][c][0.8]{$\mathsf{PEB}$  [m]}
\psfrag{Ntx}[c][c][0.8]{$\Ntx$}
\psfrag{data111111111111111111111111}[lc][lc][0.7]{Phased - Q - Averaged Orien.}
\psfrag{data2}[lc][lc][0.7]{Phased - Fixed Orien.}
\psfrag{data3}[lc][lc][0.7]{Phased - Q - Fixed Orien.}
\psfrag{data4}[lc][lc][0.7]{MIMO - Fixed Orien.}
\psfrag{data5}[lc][lc][0.7]{MIMO - Averaged Orien. }
\psfrag{data6}[lc][lc][0.7]{Timed - Averaged Orien. }
\psfrag{data7}[lc][lc][0.7]{Phased - Averaged Orien. }
\psfrag{data8}[lc][lc][0.7]{Timed - Fixed Orien.}
\psfrag{data9}[lc][lc][0.7]{Timed - Q - Averaged Orien.}
\psfrag{data10}[lc][lc][0.7]{Timed - Q - Fixed Orien.}
\centerline{\includegraphics[width=0.45\textwidth]{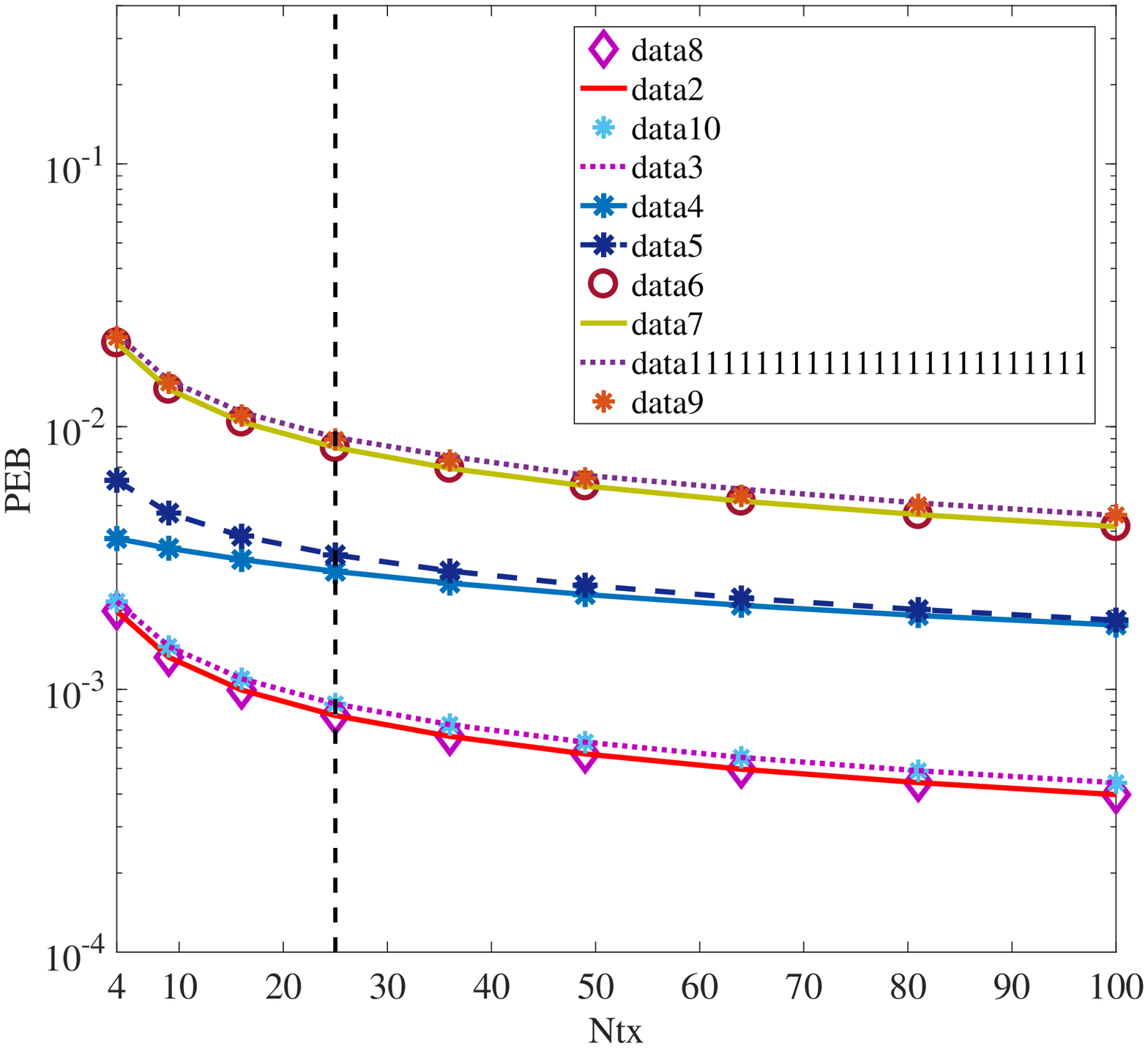}}
\caption{$\mathsf{PEB}$ \textit{vs.} $\Ntx$, $\Nrx=25$ and orientation aware.} 
\label{fig:peb_awgn_ntx} 
\end{figure}
Figure \ref{fig:peb_awgn_ntx} reports the \ac{PEB} performance as a function of $\Ntx$  {and of the {\ac{Rx}} orientation} in free-space. MIMO, timed and phased arrays  {(with and without quantization errors)} are compared in the \textit{orientation-aware} case when the number of receiving antennas is kept fixed to $\Nrx=25$. 
\begin{figure}
\psfrag{data2}[lc][lc][0.7]{Timed}
\psfrag{data3}[lc][lc][0.7]{Phased}
\psfrag{data11111111}[lc][lc][0.7]{Phased - Q }
\psfrag{data4}[lc][lc][0.7]{MIMO}
\psfrag{data5}[lc][lc][0.7]{MIMO - O}
\psfrag{data6}[lc][lc][0.7]{Random}
\psfrag{data7}[lc][lc][0.7]{Timed - Q}
\psfrag{Nrx}[lc][lc][0.8]{$\Nrx$}
\psfrag{PEB}[lc][lc][0.8]{$\mathsf{PEB}$ [m]}
\psfrag{OEB}[lc][lc][0.8]{$\mathsf{OEB}$ [deg]}
\centerline{\includegraphics[width=0.45\textwidth]{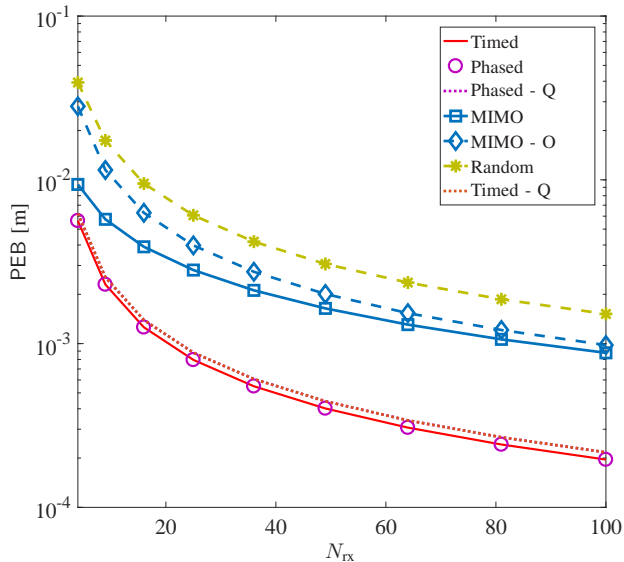}}
\caption{$\mathsf{PEB}$ \textit{vs.} $\Nrx$, $\Ntx=25$ and fixed receiver orientations.} \label{fig:ccompideal1}
\end{figure}
%

It can be observed that \ac{MIMO} arrays, relying on different transmitted waveforms, outperform timed and phased arrays  {in averaged orientations cases; whereas, in fixed {\ac{Rx}} orientation, as described in} Sec.~\ref{sec:idealscenario}, arrays  {operating beamforming} exhibit a better performance.  {This is due to the fact that} beamforming strategies (e.g., those adopted in timed and phased arrays) fail in preserving the same accuracy for any geometric configuration (i.e., for any {\ac{Rx}} orientation). Contrarily, thanks to the diversity gain characterizing \ac{MIMO} arrays, {\ac{Rx}} orientations have a less significant effect on positioning accuracy. 

{For what the beamforming arrays are concerned, it can be observed that, as expected, timed and phased arrays results coincide as  $W/\fc \ll 1$. In fact, phased arrays are the best candidate to be adopted in narrowband systems where there is no need to compensate for delays to perform beamsteering operation. We refer the reader to our previous work \cite{guerra2015position} to better appreciate the impact of the fractional bandwidth on the timed/phased array performance.} 
%
%
\begin{figure}[t!]
\psfrag{PEB}[lc][lc][0.8]{$\mathsf{PEB}$  [m]}
\psfrag{Nrx}[lc][lc][0.8]{$\Nrx$}
\psfrag{data2}[lc][lc][0.7]{MIMO}
\psfrag{data3}[lc][lc][0.7]{Timed}
\psfrag{data4}[lc][lc][0.7]{Phased}
\psfrag{data5}[lc][lc][0.7]{Timed - Q}
\psfrag{data6}[lc][lc][0.7]{Random}
\psfrag{data7}[lc][lc][0.7]{MIMO - O}
\psfrag{data8}[lc][lc][0.7]{Timed - S}
\psfrag{data9}[lc][lc][0.7]{MIMO - S}
\psfrag{data10}[lc][lc][0.7]{MIMO - S - O}
\psfrag{data1111111111}[lc][lc][0.7]{Phased - Q}
\centerline{\includegraphics[width=0.45\textwidth]{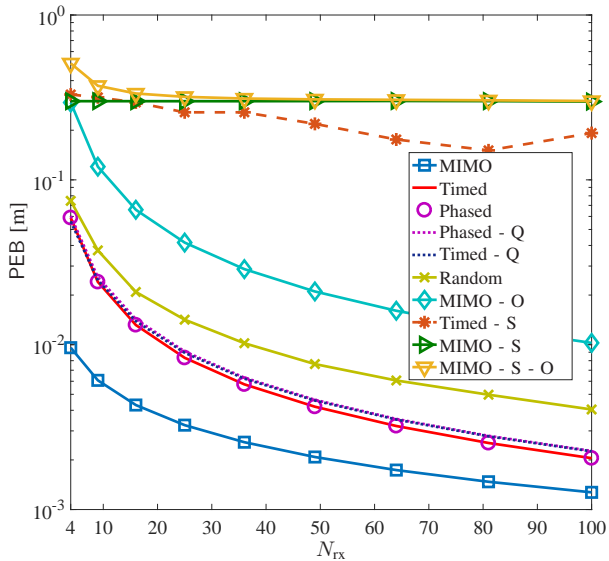}}
\caption{$\mathsf{PEB}$ \textit{vs.} $\Nrx$,  $\Ntx=25$ and averaged receiver orientation.} 
\label{fig:peb_awgn_nrx} 
\end{figure}
%
\begin{figure}[t!]
\psfrag{OEB}[c][c][0.8]{$\mathsf{OEB}$  [deg.]}
\psfrag{Nrx}[c][c][0.8]{$\Nrx$}
\psfrag{data2}[lc][lc][0.7]{MIMO - O - Averaged Orien.}
\psfrag{data3}[lc][lc][0.7]{MIMO - O - Fixed Orien.}
\psfrag{data11111111111111111111111}[lc][lc][0.75]{MIMO - S - O - Averaged Orien.}
\centerline{\includegraphics[width=0.45\textwidth]{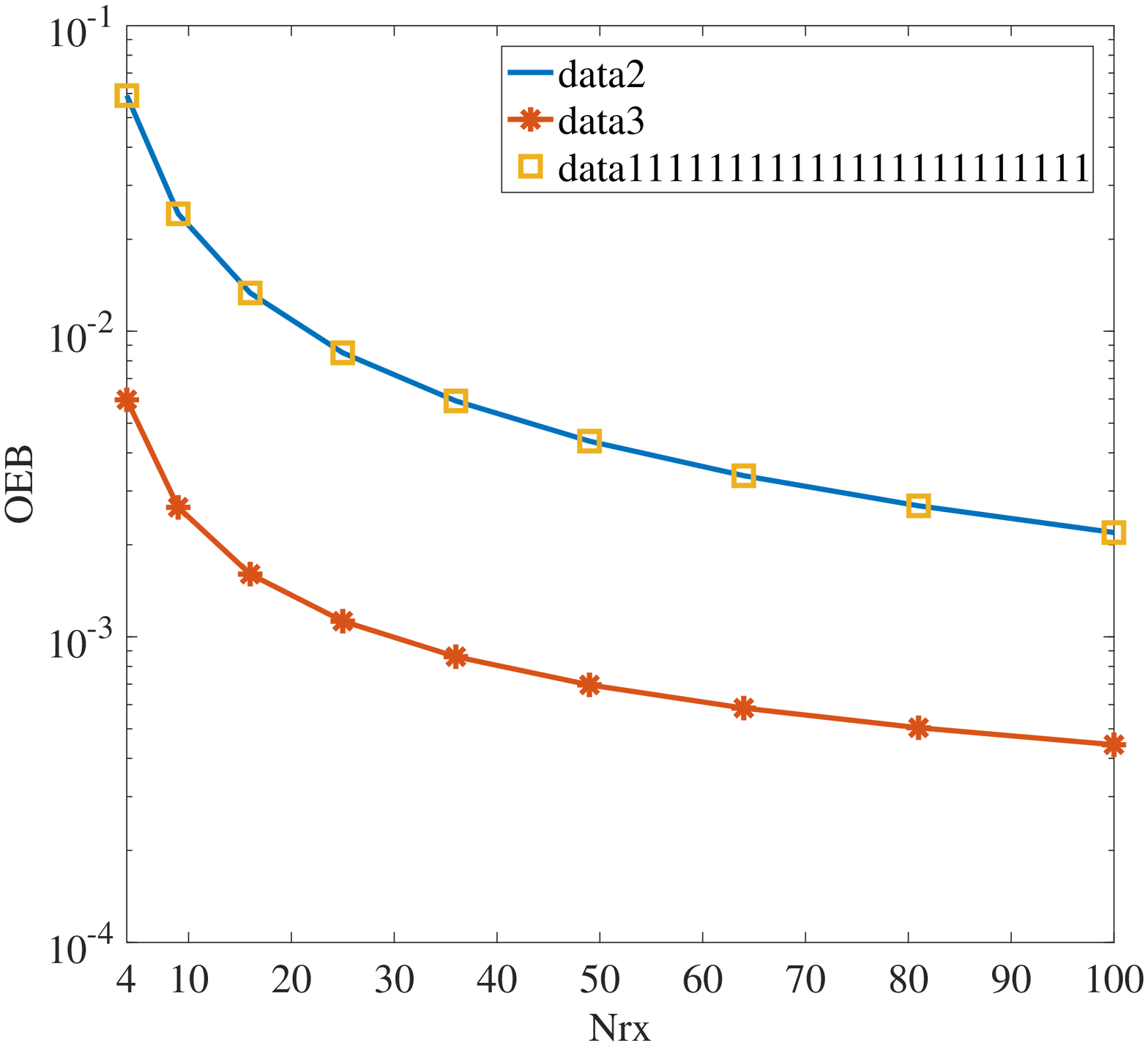}}
\caption{ $\mathsf{OEB}$ \textit{vs.} $\Nrx$,  $\Ntx=25$.} 
\label{fig:oeb_awgn_nrx} 
\end{figure}

Another important outcome from Fig.~\ref{fig:peb_awgn_ntx} is that  {array quantization errors, once characterized, slightly affect the localization performance. This implies that we can rely on simpler array structures (i.e., using switches instead of multipliers) without severely affecting the performance.} Finally, with $\Ntx \ge 25$, the performance improvement becomes less important if a sufficiently high $\Nrx $ is considered. This
implies that $\Ntx$ can be relaxed to shrink the array dimensions and to enable the integration on mobiles \cite{hong2014study}. Consequently, in the following, the number  of transmitting antennas will be fixed to $\Ntx=25$.
\paragraph{Results as a function of $\Nrx$}
\label{sec:Nrxresults}

In Fig.~\ref{fig:ccompideal1}, the \ac{PEB} performance are reported for both \textit{orientation-aware} and \textit{-unaware} cases  as a function of $\Nrx$ in free-space propagation condition, $\Ntx=25$ and fixed {\ac{Tx}} and {\ac{Rx}} orientation $\bm{\vartheta}^\tx=\bm{\vartheta}^\rx=\left[0,0 \right]^{\scriptscriptstyle \text{T}}$. The results have been obtained using the analytic expressions \eqref{eq:crbMIMOpeo}, \eqref{eq:CRBmimoknown} and \eqref{eq:crbP} and reveal that arrays performing beamforming outperform the performance of \ac{MIMO} for the particular steering and geometric configuration conditions chosen, as already observed in Fig.~\ref{fig:peb_awgn_ntx}. We ascribe the effect to an increased \ac{SNR} in the considered direction. Nevertheless, with  {arrays operating {single-beam} beamforming}, orientation estimation is not always possible and consequently the \ac{FIM} results to be singular. Matrix singularity or ill-conditioning are, here, synonymous of  the impossibility to estimate the position/orientation given the collected measurements.

Figure \ref{fig:peb_awgn_nrx} shows the average \ac{PEB} performance when the {\ac{Rx}} orientation randomly changes at each Monte Carlo iteration. For this analysis, we consider also {random weighting}, quantization errors as well as {time synchronization} mismatch between the {\ac{Tx}} and the {\ac{Rx}}. The positioning performance are shown for the \textit{orientation-aware} case if not otherwise indicated. 
\begin{figure}[t!]
\psfrag{x}[c][c][0.75]{$x$  [m]}
\psfrag{y}[c][c][0.75]{$y$  [m]}
\centerline{\includegraphics[width=0.4\textwidth]{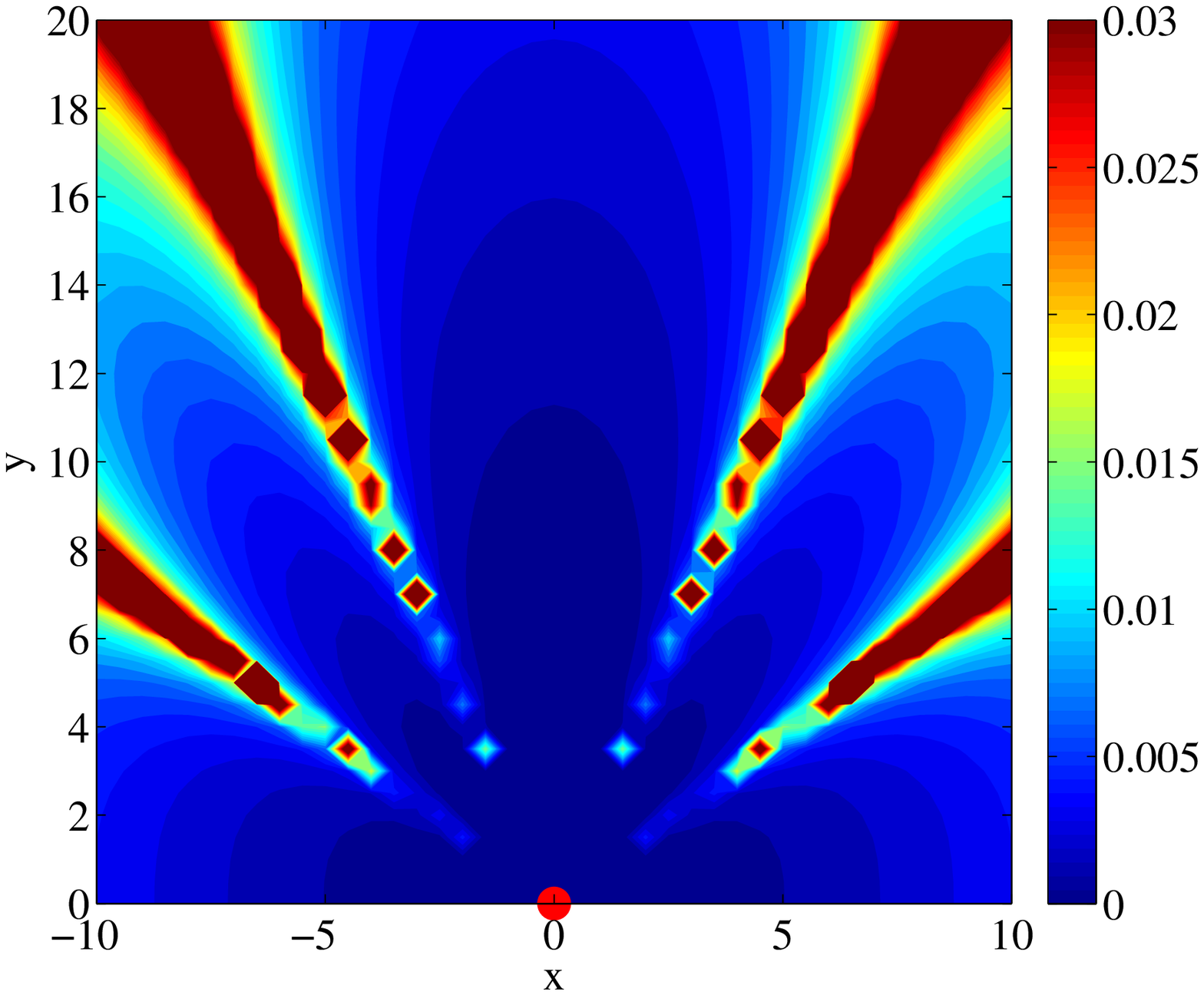}}\centerline{\includegraphics[width=0.4\textwidth]{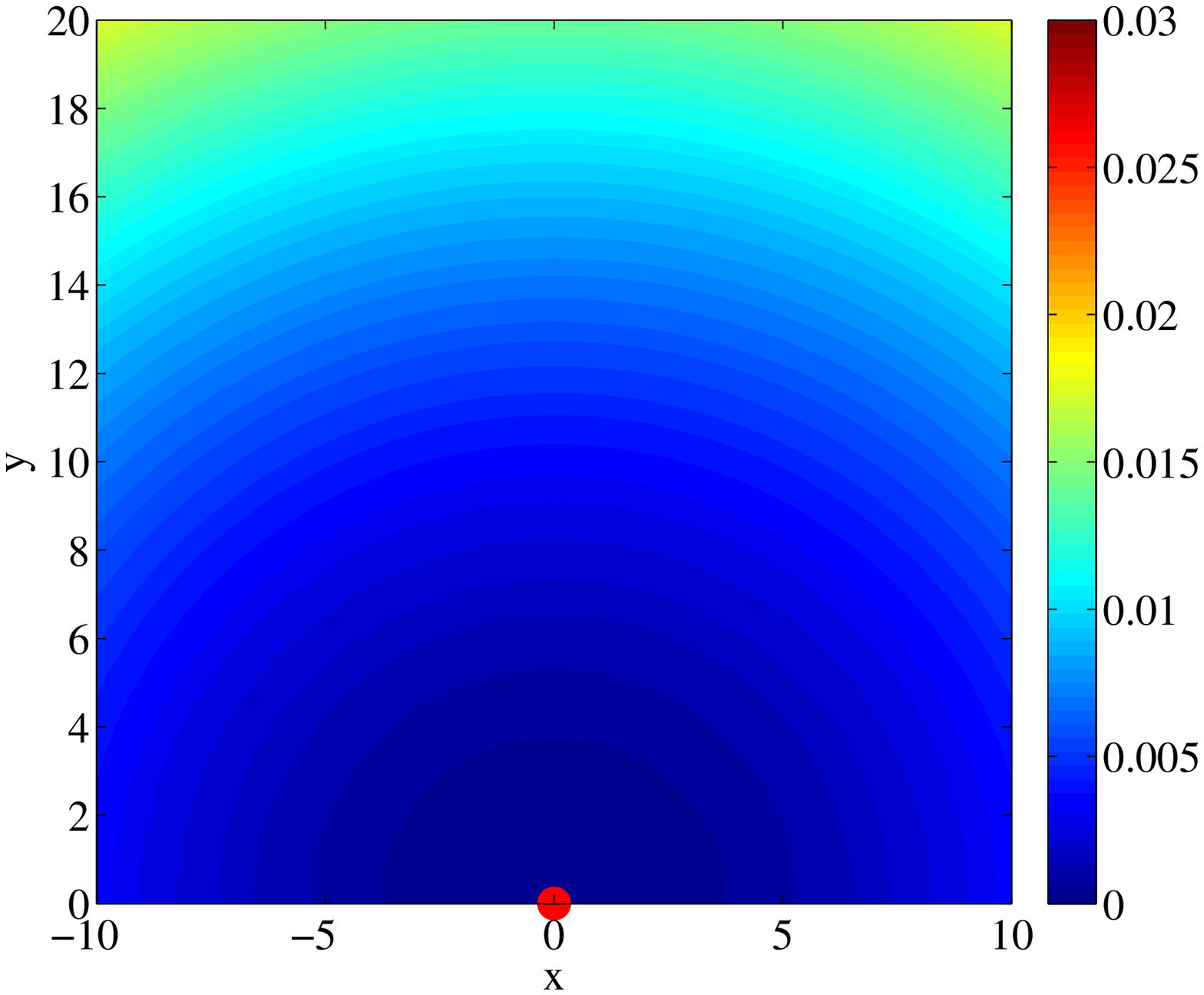}}
\caption{Phased (top) and MIMO (bottom) array $\mathsf{PEB}$,  $\Nrx=100$, $\Ntx=25$ and averaged receiver orientations.} 
\label{fig:peb_awgn_grid_phasedmimo} 
\end{figure}
As in Figs.~\ref{fig:peb_awgn_ntx}-\ref{fig:ccompideal1}, the performance of timed and phased arrays coincide due to the narrow fractional bandwidth  {(i.e., $W/\fc\approx 0.016$)}.  {Differently from Fig.~\ref{fig:ccompideal1}, in Fig.~\ref{fig:peb_awgn_nrx} \ac{MIMO} achieves a higher positioning accuracy with respect to arrays employing beamforming strategies due to the fact that results are averaged over different {\ac{Rx}} orientations. In fact, with \ac{MIMO}, a reduction in the received \ac{SNR} is experienced, but the number of independent measurements is maximized (i.e., $\Ntx\Nrx$).}

For what {random weighting} arrays are regarded, they share the structure simplicity of phased arrays but they neither perform beamforming nor achieve the diversity gain of MIMO, and thus, the positioning accuracy results degraded with respect to other structures. Nevertheless, if the localization accuracy required by the application of interest is not so stringent, they could be an interesting option to guarantee both a sub-centimeter positioning accuracy  {(e.g., for $\Nrx=50$ and $\Ntx=25$, $\mathsf{PEB} \approx 7\,$mm)} and an easy implementation in future mobiles and AP operating at \ac{mm-wave} frequencies. 

Note that when the {\ac{Tx}} orientation is one of the parameters to be estimated (\textit{orientation-unaware} case), only \ac{MIMO} results in a non-singular \ac{FIM}. Obviously, in this case, given the reduced information available at the {\ac{Rx}} side, the positioning accuracy worsen with respect to the \textit{orientation-aware} case. In all cases, the residual {time synchronization} error degrades the localization performance. 

\begin{figure}[t!]
\psfrag{PEB}[lc][lc][0.8]{$\mathsf{PEB}$ [m]}
\psfrag{Nrx}[lc][lc][0.8]{$\Nrx$}
\psfrag{OEB}[lc][lc][0.8]{$\mathsf{OEB}$ [deg]}
\psfrag{data1}[lc][lc][0.7]{\quad $L=1$}
\psfrag{data2}[lc][lc][0.7]{\quad $L=2$}
\psfrag{data666666666}[lc][lc][0.7]{\quad $L=5$}
\centerline{\includegraphics[width=0.4\textwidth]{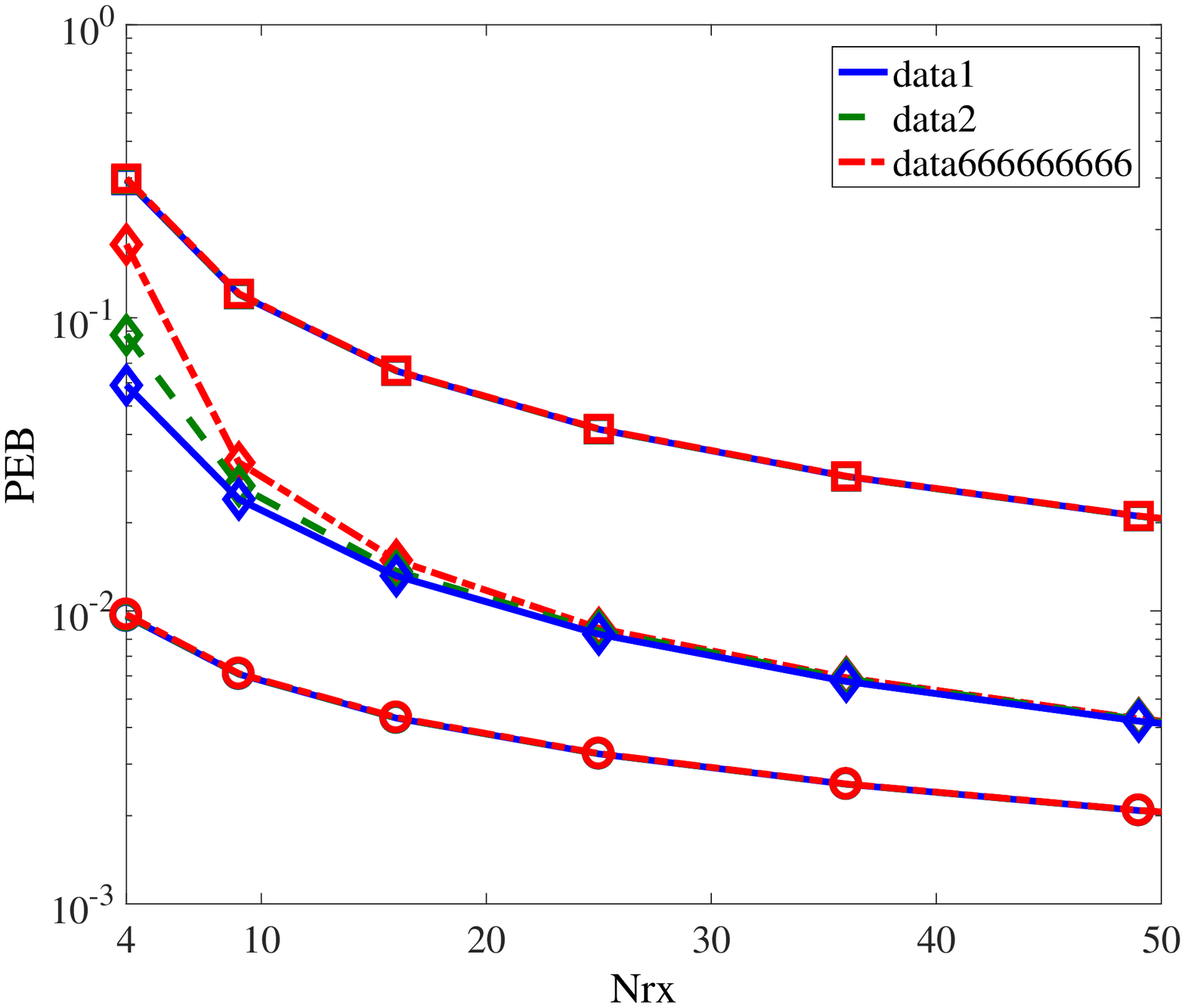}}
\psfrag{data1}[lc][lc][0.7]{\quad $L=1$}
\psfrag{data2}[lc][lc][0.7]{\quad $L=2$}
\psfrag{data33333333}[lc][lc][0.7]{\quad $L=5$}
\centerline{\includegraphics[width=0.4\textwidth]{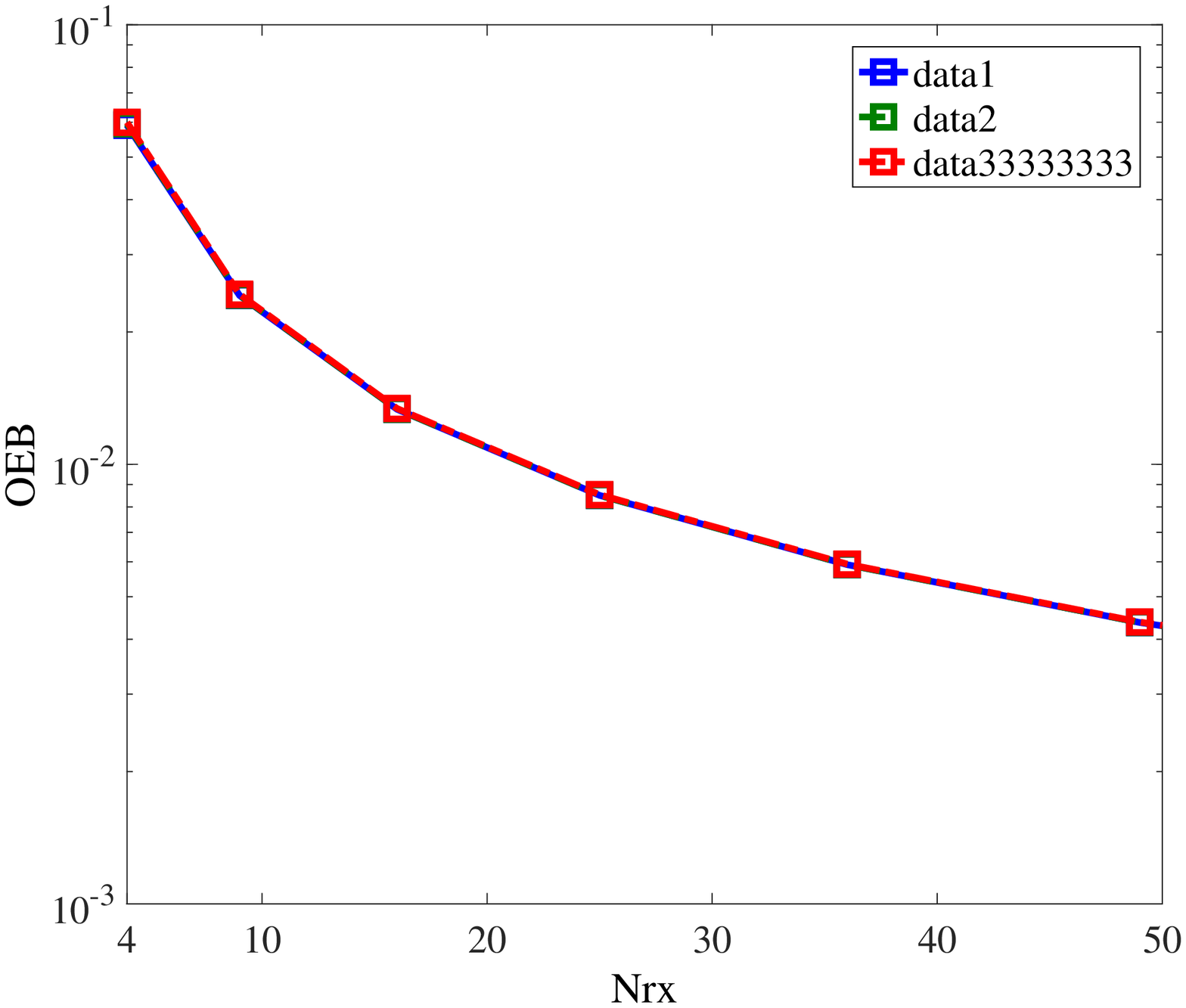}}
\caption{\ac{PEB} and \ac{OEB} \textit{vs.} $\Nrx$ in multipath propagation scenario, $\Ntx=25$, $W=1\,$GHz, averaged receiver orientation. Diamond marked lines refer to phased array, circle marked lines to \ac{MIMO}, and square marked lines to \ac{MIMO} orientation unaware.} 
\label{fig:peb_oeb_MP} 
\end{figure}
 {Figure \ref{fig:oeb_awgn_nrx} reports the \ac{OEB} as a function of $\Nrx$. In this case, only the performance of \ac{MIMO} is present because of the singularity problem arising in timed, phased and {random weighting} arrays. An interesting results is that, in this case, {time synchronization} error does not impact the orientation accuracy.}
\noindent \paragraph{Grid results}
\label{sec:Gridresults}

Figure~\ref{fig:peb_awgn_grid_phasedmimo} reports the \ac{PEB} results for the \textit{orientation-aware} case when the mobile moves in a grid of points spaced of $0.5\,$m considering phased (Fig.~\ref{fig:peb_awgn_grid_phasedmimo}-top) and \ac{MIMO} (Fig.~\ref{fig:peb_awgn_grid_phasedmimo}-bottom) arrays, respectively. We considered a 3D indoor scenario of $10\times 10 \times 3\,$m$^3$ where the mobile and the \ac{AP} centroids are at the same height. The {\ac{Rx}}, equipped with $\Nrx=100$ antennas, is placed in $\pr=\left[ 0,0,0\right]^{\scriptscriptstyle \text{T}}$ with orientation changing at each Monte Carlo iteration. On the other side, the mobile array is equipped with $\Ntx=25$ and its orientation and its steering direction are fixed to $\orien=\left[0,0 \right]^{\scriptscriptstyle \text{T}}$ and to the broadside direction, respectively. 
Grid results confirm that \ac{MIMO} arrays localization performance does not depend on the {\ac{Rx}} orientation and on mobile position in space but only on the distance between the {\ac{Tx}} and the {\ac{Rx}}. Indeed, when comparing Fig.~\ref{fig:peb_awgn_grid_phasedmimo}-top and Fig.~\ref{fig:peb_awgn_grid_phasedmimo}-bottom, it can be seen that if the mobile steering is fixed, the localization accuracy is higher in a privileged direction in space corresponding to the best geometric configuration conditions. 

\subsubsection{Multipath scenario}
\label{sec:ch3_MPresults}

\paragraph{Results as a function of $\Nrx$}
\label{sec:Nrxresults_MP}

{Figure~\ref{fig:peb_oeb_MP} investigates the multipath effect by analysing the \ac{PEB} and \ac{OEB} averaged over different {\ac{Rx}} orientations and as a function of the number of \acp{MPC} for phased and \ac{MIMO} arrays.} We consider the statistical multipath parameters described in Sec.~\ref{sec:ch3_sysconfig}.
As foreseen in the asymptotic analysis in Sec.~\ref{sec:mp_loc_ac}, when increasing $\Nrx$, the \acp{MPC} effect becomes negligible and the performance tends to coincide with that obtained in free-space.  Moreover, it is possible to remark that phased arrays are more sensitive to multipath with respect to MIMO arrays, especially when the number of receiving antennas is small.  {In fact, for phased arrays at least $\Ntx=25$ and $\Nrx=25$ are necessary to make the \acp{MPC} fully resolvable.  {In any case, it is interesting to note that to make the impact of \acp{MPC} negligible the antenna arrays are not required to be significantly massive.}}
\vspace{-0.3cm}
\section{Conclusion}
\label{sec:conclusions}

In this paper, we have considered a new scenario where a single-anchor localization exploiting \ac{mm-wave} massive arrays has been put forth for next 5G applications. The theoretical localization performance has been evaluated by deriving the position and orientation \ac{CRB} for different array configurations. Phase quantization, the residual {time synchronization} mismatch, and multipath have been considered as nuisance parameters. The comparison between \ac{MIMO} and beamforming has been carried out by analyzing the impact of the number of antenna elements on the performance. 
 {From {analytical and} simulation results, the main conclusions and achievements emerged in this paper can be summarized as follows:
 {
\begin{itemize}
\item {We show through an asymptotic analysis (i.e., $\Nrx \rightarrow \infty$) that the considered \ac{CRB} is a tight bound regardless the propagation condition and the array configuration considered.}
\item A beamforming with steering capability is desirable to maximize the \ac{SNR} towards a specific direction, as happens for phased and timed arrays; but it reduces the diversity gain (typical of \ac{MIMO}) exploitable for retrieving positioning information, especially in \textit{orientation-unaware} situations. 
\item Quantization errors slightly impact the localization performance of both timed and phased arrays. Consequently, the array design requirements can be relaxed in favor of lower complexity and cost.
\item Both for MIMO and beamforming, a {time synchronization} mismatch between the {\ac{Tx}} and the {\ac{Rx}} significantly degrades the positioning performance. 
From this point-of-view, it is important to minimize the clock mismatch between the arrays. Contrarily, in the \textit{orientation-unaware} case, the {time synchronization} error does not affect the performance of the estimation;
\item  The adoption of multiple antennas makes the positioning insensitive to multipath for most of geometric configurations. This is true even when the number of antennas is not extremely high (i.e., $\Ntx, \Nrx >20$); {we demonstrated such point also through an asymptotic analysis.}
\item Finally, {random weighting} results to be a good low-complexity strategy to achieve centimeter level accuracy.
\end{itemize}
}}

\appendices
\section{}\label{appA}

Considering the sub-set of random parameters  {$ \bm{\psi}_\text{r} =\left[\bm{\kappa}_2^{\scriptscriptstyle \text{T}},\, \ldots,\, \bm{\kappa}_L^{\scriptscriptstyle \text{T}} ,\, \sincro \right]^{\scriptscriptstyle \text{T}}  =\left[\bm{\alpha}^{\Re\,\scriptscriptstyle \text{T}},\, \bm{\alpha}^{\Im\,\scriptscriptstyle \text{T}} ,\, \sincro   \right]^{\scriptscriptstyle \text{T}}$}, and treating them as independent \acp{RV} we can write
\begin{align}
&\ln \left(f(\bm{\psi}_\text{r})\right)=\ln\left(f(\bm{\alpha}^{\Re \,\scriptscriptstyle \text{T}})\right)+\ln\left(f(\bm{\alpha}^{\Im \,\scriptscriptstyle \text{T}})\right)+\ln\left(f(\sincro)\right).
\end{align}
Therefore, the elements of the \textit{a-priori} \ac{FIM} are

\begin{align}
&J_{\sincro \sincro}^{\text{p}}=\frac{1}{\varsincro},  \quad J_{\alpha^{\Re}_k \alpha^{\Re}_l}^{\text{p}}=J_{\alpha^{\Im}_k \alpha^{\Im}_l}^{\text{p}}=\begin{cases}
& \frac{1}{\sigma_l^2}\,\,\,\text{if $l = k$}\\
&0\quad\,\text{if $l \neq k$}
\end{cases}.
\end{align}
\vspace{-0.3cm}
\section{}\label{appB}
\newcounter{MYtempeqncnt3}
\begin{figure*}[t!]
\normalsize
\setcounter{MYtempeqncnt3}{\value{equation}}
\begin{align}\label{eq:Gmatrix_part} 
&{\nabla_{\mathbf{q}\mathbf{q}}\!\left( \tau_{im1}, \tau_{jm1} \right)}\!\!=\!\!\frac{d_\text{ant}^2}{(c\,y)^2} \left[\begin{array}{lllll}
\! \left(m_x-i_x\right)\left(m_x-j_x\right)\!&\frac{y}{d_\text{ant}}(m_x-i_x)\!&\!(m_x-i_x)(j_z-m_z)\!& y\,(i_x-m_x)\,j_z\hdots \!&\!y\,(i_x-m_x)\,j_x \\
\hdots  \!&\!\frac{y^2}{d_\text{ant}^2}\!&\!\frac{y}{d_\text{ant}} (j_z-m_z)\!& \! -\frac{y^2}{d_\text{ant}} j_z\!&\! -\frac{y^2}{d_\text{ant}} j_x \\
\hdots   &\hdots  &\left(i_z-m_z\right)\left(j_z-m_z\right)& \! y\,(m_z-i_z)\,j_z&\!y\,(m_z-i_z)\,j_x\\
\hdots   &\hdots  &\hdots  & \! {y^2}i_z j_z\!&\!{y^2}i_z j_x \\
\hdots   &\hdots  &\hdots  &\hdots  &{y^2}i_x j_x\\
\end{array} \right] 
\end{align}
\hrulefill
\begin{align} \label{eq:Gresults1}
&\mathsf{CRB}\left(\mathbf{q} \right)\!\!=\!\! \frac{c^2}{8 \pi^2\,\Ntx\, \mathsf{SNR}_1 \left(\beta^2_i+\fc^2\right)} \frac{1}{S} \left[\begin{array}{lllll}
\! \frac{12}{(\Nrx -1)}\!&0\!&\!0\!& 0\!&\!  \frac{12}{ y (1-\Nrx)}\\
0\!&\!\frac{S}{\Nrx}\!&\!0\!& \! 0\!&\!0 \\
0&0&\frac{12}{(\Nrx-1)}& \! \frac{12}{ y\, (1-\Nrx)}&\!0\\
0&0&\frac{12}{ y\, (1-\Nrx)}& \!   \frac{12 \,(\Nrx +\Ntx-2)}{y^2  \,(\Ntx - 1)\,(\Nrx - 1)}\!&\!0 \\
\frac{12}{ y (1-\Nrx)}&0&0&0&\frac{12 \,(\Nrx +\Ntx-2)}{y^2  \,(\Ntx - 1)\,(\Nrx - 1)}\\
\end{array} \right]\\ \label{eq:Gresults2}
&\mathsf{CRB}\left(\mathbf{p} {^\tx} \right)\!\!=\!\! \frac{c^2}{8 \pi^2\, \Ntx\,\mathsf{SNR}_1  \left(\beta^2+\fc^2\right)}\frac{1}{S} \, \text{diag}\left( \frac{12}{\Ntx(\Nrx-1)},\frac{S}{\Ntx\,\Nrx}, \frac{12}{\Ntx (\Nrx-1)}  \right) 
\end{align}
\hrulefill
\vspace*{4pt}
\end{figure*} 
In this Appendix we derive the elements of the data \ac{FIM} reported in \eqref{eq:jtheta}. To accomplish this task, we introduce the following quantities
\begin{align}
&\chi_{ij}^{(l,k)}(p)\!\!=\!\!\int_W\!\!\!\tilde{b}_{ij}(f)\left(f+\fc \right)^p\! e^{-j\, 2\, \pi\, f\, \Delta \tau_{ij}^{(l,k)}}P_i(f)\, P_j^*(f)df    \nonumber \\
&{R^p_{ij}(\Delta \tau)}=\int_W\,\tilde{b}_{ij}(f)\, e^{-j\, 2\, \pi\, f\, \Delta \tau }\,P_i(f)\, P_j^*(f)\, df  \nonumber \\
& R^{\ddot{p}}_{ij}(\Delta \tau)=\int_W\,\tilde{b}_{ij}(f)\,f^2\, e^{-j\, 2\, \pi\, f\, \Delta \tau }\,P_i(f)\, P_j^*(f)\, df  
\end{align}
where $\Delta \tau_{ij}^{(l,k)}=\tau_{iml}-\tau_{jmk}$, $\tilde{b}_{ij}(f)=\tilde{b}_i(f)\, \tilde{b}_j^*(f)$, and $\tilde{b}_{ij}^\text{c}=\tilde{b}_i^\text{c}\, \left(\tilde{b}_j^\text{c}\right)^*$.  {The elements of $\mathbf{J}_{\mathbf{q}\mathbf{q}}^{\text{d}}$ can be expressed as in \eqref{eq:Jqq_main}.} The elements of $\FIMealfalf^{\text{d}}$ are
\begin{align*}
&J_{ a_1 \, a_1}\!=\!2\,\nu \sum_{mij}\Re \left\{ \tilde{b}_{ij}^\text{c} \,  \xi_{ij}^{(1,1)}\,  R^p_{ij} \left( \Delta \tau_{ij}^{(1,1)} \right) \right\}  
\end{align*}
\begin{align*}
&J_{\alpha_k^\Re\, a_1}\!=\!J_{a_1\,\alpha_k^\Re}^{\scriptscriptstyle \text{H}}= 2\,\nu  \sum_{mij}\,\Re\left\{ \tilde{b}_{ij}^\text{c} \,  \xi_{ij}^{(1,k)}\,  R^p_{ij} \left( \Delta \tau_{ij}^{(1,k)} \right) \right\}  
\end{align*}
\begin{align*}
&J_{\alpha_k^\Im \, a_1}\!=\!J_{a_1 \,\alpha_k^\Im}^{\scriptscriptstyle \text{H}}=2\,\nu \sum_{mij}\,\Im \left\{ \tilde{b}_{ij}^\text{c} \,  \xi_{ij}^{(1,k)}\,  R^p_{ij}\left( \Delta \tau_{ij}^{(1,k)}\right)  \right\} 
\end{align*}
\begin{align*}
&J_{ \alpha_k^\Re \, \alpha_l^\Re}\!=\!J_{ \alpha_k^\Im \, \alpha_l^\Im}\!=\!2\,\nu \sum_{mij}\Re \left\{ \,\tilde{b}_{ij}^\text{c} \,  \xi_{ij}^{(l,k)} \,  R^p_{ij}\left( \Delta \tau_{ij}^{(l,k)} \right) \right\}  
\end{align*}
\begin{align*}
&J_{ \alpha_k^\Im \, \alpha_l^\Re}\!=\!J_{ \alpha_l^\Re\,\alpha_k^\Im }^{\scriptscriptstyle \text{H}}=2\,\nu \sum_{mij}\Im \left\{\,\tilde{b}_{ij}^\text{c} \,  \xi_{ij}^{(l,k)} \,  R^p_{ij}\left( \Delta \tau_{ij}^{(l,k)} \right)  \right\} 
\end{align*}
where $\xi_{ij}^{(1,k)}=e^{-j\, 2\, \pi\,\fc \left( \tau_{im1}+ \sincro+\tau_{m}^\rx({\bm{\theta}_k})-\tau_{j}^\tx({\bm{\theta}_k}) \right)}$ and  $\xi_{ij}^{(l,k)}=e^{-j\, 2\, \pi\, \fc \left(-\tau_m^\rx({\bm{\theta}_l})+\tau_m^\rx({\bm{\theta}_k})+\tau_i^\tx({\bm{\theta}_l})-\tau_j^\tx({\bm{\theta}_k}) \right)}$. 

\noindent The elements of $\mathbf{J}_{\mathbf{q}\bm{\kappa}}^{\text{d}}$ are 
\begin{align}\label{eq:Jqkappa}
&J_{q\,a_1 }= 4\, \pi\, a_1 \,\nu \sum_{mij}\Im\left\{\,\tilde{b}_{ij}^\text{c} \,\xi_{ij}^{(1,1)}\,  \chi_{ij}^{(1,1)}(1)  \right\}\nabla_{q}\left( \tau_{im1}\right)=0 \nonumber \\
&J_{q \, \alpha_k^\Re}=- 4\, \pi\, a_1 \,\nu  \sum_{mij}\Im \left\{\,\tilde{b}_{ij}^\text{c} \,  \xi_{ij}^{(k,1)} \chi_{ij}^{(k,1)}(1)  \right\}\nabla_{q}\left( \tau_{jm1}\right) \nonumber \\
&J_{q \, \alpha_k^\Im}= 4\, \pi\, a_1 \,\nu \sum_{mij}\Re \left\{\, \tilde{b}_{ij}^\text{c} \,\xi_{ij}^{(k,1)} \chi_{ij}^{(k,1)}(1)  \right\}\nabla_{q}\left( \tau_{im1}\right)  
\end{align}
where $\xi_{ij}^{(k,1)}=e^{-j\, 2\, \pi\, \fc\, \left( -\tau_{jm1}-\sincro-\tau_{m}^\rx({\bm{\theta}_k})+\tau_{i}^\tx({\bm{\theta_k}}) \right)}$. Note that $\mathbf{J}_{\bm{\kappa}\mathbf{q}}^{\text{d}}=\mathbf{J}_{\mathbf{q}\bm{\kappa}}^{\text{d}\, \scriptscriptstyle \text{H}}$ .

\noindent Now, if we consider the presence of a residual {time synchronization} error, the \ac{FIM} $\mathbf{J}_{\sincro\, \sincro}^\text{d}$ is
\begin{align*}
J_{\sincro\,\sincro }&= 8\, \pi^2 \, \nu \,\Re\left\{\sum_{mij}\,  \tilde{b}_{ij}^\text{c} \left[a_1^2 \, \xi_{ij}^{(1,1)}\, \chi_{ij}^{(1,1)}\left(2 \right) +\right. \right. \nonumber \\
&\left.\left. + \sum_{k=2}^{L} \sigma_l^2\, R^{\ddot{p}}_{ij}\left(\Delta \tau_{ij}^{(k,k)} \right)\, e^{-j\, 2\, \pi\, \fc \, \left(\tau_i^\tx ({\bm{\theta}_k})-\tau_j^\tx ({\bm{\theta}_k}) \right)} \right] \right\}.
\end{align*}
The elements of $\mathbf{J}_{\bm{\kappa} \sincro}^\text{d}$  are
\begin{align*}
&J_{ a_1 \, \sincro}= 4 \, \pi\, a_1\, \nu\, \sum_{mij}\,\,\Im\left\{\tilde{b}_{ij}^\text{c} \, \xi_{ij}^{(1,1)}\,\chi_{ij}^{(1,1)}\left(1 \right) \right\} \nonumber  \\
&J_{ \alpha_k^\Re \,\sincro }= 4 \, \pi\, a_1\, \nu\, \sum_{mij}\,\Im\left\{ \tilde{b}_{ij}^\text{c}\, \xi_{ij}^{(1,k)}\,\chi_{ij}^{(1,k)} \left(1 \right)  \right\} \nonumber \\
&J_{\alpha_k^\Im \,\sincro }= 4\, \pi\, a_1 \, \nu\,\sum_{mij}\,\Re\left\{\tilde{b}_{ij}^\text{c}\, \xi_{ij}^{(1,k)}\, \chi_{ij}^{(1,k)}\left(1 \right) \right\}.
\end{align*}
As before the elements of $\mathbf{J}_{\sincro\, \bm{\kappa}}^\text{d}$ could be found as $\mathbf{J}_{\sincro\, \bm{\kappa}}^\text{d}=\mathbf{J}_{\bm{\kappa} \sincro}^{\text{d}\, \scriptscriptstyle \text{H}}$

\noindent Finally, the elements of $\mathbf{J}_{\mathbf{q}\sincro}^{\text{d}}$ are
\begin{align*}
&J_{q \sincro}= 8 \, \pi^2 \, \nu\, a_1^2\, \sum_{mij}\Re\left\{\,\tilde{b}_{ij}^\text{c}\, \xi_{ij}^{(1,1)}\, \chi_{ij}^{(1,1)}(2) \right\} \nabla_{q} \left( \tau_{jm1}\right)  
\end{align*}
and $\mathbf{J}_{\sincro\mathbf{q}}^{\text{d}}=\mathbf{J}_{\mathbf{q}\sincro}^{\text{d}\, \scriptscriptstyle \text{H}}$ .

\vspace{-0.3cm}
\section{}\label{appC}
In  this Appendix  we will specialize the expression of the symmetric matrix $\mathbf{G}$ reported in \eqref{eq:crbawgn}-\eqref{eq:crbawgn_elem}.  {To this end, we explicit the geometric relationship relating the \ac{TOA} between each TX-RX antennas couple and the considered localization (position or orientation) parameter, i.e., $\nabla_{q_a \, q_b}\!\left(\tau_{iml},\tau_{jml}\right)=\nabla_{q_a}\!\left(\tau_{iml}\right)\!\nabla_{q_b}\!\left(\tau_{jml}\right)$}. 
For the particular antenna configuration chosen described in Sec.~\ref{sec:planar}, in which the array antennas are spaced of $d_\text{ant}$, and considering  {$\orien=\left[0, 0 \right]^{\scriptscriptstyle \text{T}}$}, we can  {compute a simplified version of \eqref{eq:der_toa_q0}-\eqref{eq:nablaorie}. Specifically, it is possible to obtain:
\begin{align}\label{eq:der_toa_p_planar}
&\nabla_p\left( \tau_{im1} \right)=\frac{1}{c} \left[c\,\nabla_p\left( \tau_{1} \right)+d_\text{ant} \left((i_x-m_x) \nabla_p(\phi_1) \right. \right. \nonumber \\
&\left. \left. \qquad\qquad + (m_z-i_z) \nabla_p(\theta_1) \right) \right] \\ \label{eq:der_toa_o_planar}
&\nabla_{\vartheta^\tx}\left( \tau_{im1} \right)=-\frac{d_\text{ant}}{c}\, i_z \, , \,\,\,\nabla_{\varphi^\tx}\left( \tau_{im1} \right)=-\frac{d_\text{ant}}{c}\, i_x
\end{align}
}
with $m_x=m_z=-\frac{\sqrt{\Nrx}-1 }{2},-\frac{\sqrt{\Nrx}-1}{2}+1, \ldots, \frac{\sqrt{\Nrx}-1}{2}$ and $i_x=i_z=j_x=j_z=-\frac{\sqrt{\Ntx}-1 }{2},-\frac{\sqrt{\Ntx}-1}{2}+1, \ldots, \frac{\sqrt{\Ntx}-1}{2}$.
 {From (\ref{eq:der_toa_p_planar})-(\ref{eq:der_toa_o_planar}), it is straightforward to derive (\ref{eq:Gmatrix_part}). Then,} by considering the summations  {present in $\mathbf{G}$}, it is possible to obtain the \ac{CRB} matrices for \ac{MIMO} and timed arrays respectively as in \eqref{eq:Gresults1}-\eqref{eq:Gresults2} where $S=A^\rx/y^2$.

{\section{}\label{appD}
In this Appendix we consider the \ac{AF} as defined in \eqref{eq:AFnorm}. \\
\indent The \ac{AF} for the position coordinates shows a main peak in correspondence of the true \ac{Tx} position and secondary sidelobes peaks relative to ``wrong" positions. An ambiguity problem arises when one of these sidelobes overcomes or becomes comparable to the main beam due to noise. Consequently, to determine whether ambiguities are negligible in the non-massive array case, we have derived a threshold on the noise standard deviation in order to keep the ambiguity probability fixed to a desired low value. \\
\indent By comparing the threshold obtained with the value used in the numerical results, we can demonstrate that we operate at a high \ac{SNR} regime where the \ac{CRB} is tight even if a non-massive array is adopted.
\\
\begin{table}[t!]
\caption{{MIMO \textit{vs.} beamforming comparison}}
\label{tab:tab1}
\begin{center}
{
\begin{tabular}{|l|l|l|l|l|l|l|l|}
\hline &&&& \\
MIMO/Timed & $\Nrx$ & $\gamma$ [dB]& $\sigma_\text{thr}$ [mV] &   $\sigma_\text{sim}$ [mV]  \\ &&&& \\ \hline &&&& \\ 
MIMO & 4   &  -36.9&$0.062$ &$0.022$ \\&&&& \\  
MIMO & 36 &    -32.1& $0.187$& $0.022$ \\ &&&& \\ 
MIMO & 100  &-29.9 &$0.313$ & $0.022$ \\&&&& \\  
Phased & 4 & -33.5&$0.136$ & $0.022$   \\&&&& \\  
Phased & 36 & -28.7 &$0.406$ &$0.022$    \\&&&& \\  
Phased & 100  & -26.5& $0.677$& $0.022$  \\ &&&& \\ 
\hline
\end{tabular}
}
\end{center}
\end{table}
To this end, we define the ambiguity probability as \cite{van2004detection}
\begin{align}
&{\text{P}_\text{A}=   \frac{1}{2}\, \text{erfc}\left( \frac{\gamma}{\sqrt{4\,\sigma^2}} \right)}
\end{align}
where $\sigma$ is the noise standard deviation and $\gamma$ is the gap between the main lobe of the \ac{AF} and the highest secondary sidelobe. Then, given a certain gap $\gamma$, it is possible to compute the noise threshold as
\begin{align}
&{\sigma_\text{thr}=\frac{\gamma}{{2}}\, \frac{1}{\text{erfc}^{-1}\left({2\, \text{P}_\text{A}^*} \right)} }\,.
\end{align}
In Table~\ref{tab:tab1}, we report the obtained simulation results. We have considered the \ac{Tx} moving in a grid of points spaced apart of $0.2\,$m in a cube of dimension $8\times8\times8\,$m$^3$. The target ambiguity probability has been fixed to $10^{-2}$. The gap $\gamma$ has been set to the minimum side-lobe level considering the three spatial coordinates (that is, to the worst case scenario). $\sigma_\text{sim}$ represents the noise standard deviation used in the numerical results of the paper.\\
\indent As one can notice, in all the tested configurations the noise standard deviation used in the numerical results is always much lower than the threshold $\sigma_\text{thr}$ above that the ambiguity effect is not anymore negligible. \\
\indent The proposed method is a useful tool to test whether a specific scenario can be considered in the asymptotic regime and, hence, the \ac{CRB} can be a meaningful metric.
}



\end{document}